\documentstyle[12pt,amssymbols]{article}
\input{epsf}

\newcommand{\beq}{\begin{equation}}
\newcommand{\eeq}{\end{equation}}
\newcommand{\mnod}{m_Q^{(0)}}
\begin{document}
\begin{titlepage}

\begin{flushright}
{\small OHSTPY-HEP-T-97-006\\
GUTPA/97/4/1\\
UCSD/PTH/97-10\\
UK/97-07 \\
\today}
\end{flushright}

\vspace*{3mm}

\begin{center}
{\LARGE \bf Heavy-Light  Mesons with \\
Quenched Lattice NRQCD: \\
 Results on Decay Constants}\\[12mm]

{ \large  A.~Ali Khan\footnote{UKQCD Collaboration}${\vphantom{\Large 
A}}^{,}$\footnote{former address: University of Glasgow, Glasgow
G12 8QQ, U.K.} 
and J.~Shigemitsu}\\
The Ohio State University \\
Columbus, OH 43210-1106, USA \\
\vspace{3mm}
{\large  S. Collins${\vphantom{\Large A}}^{1,}$\footnote{former address:
SCRI, Florida State University, Tallahassee, FL 32306, USA} and 
C.~T.~H.~Davies$
{\vphantom{\Large A}}^1$} \\University of Glasgow, \\
Glasgow G12~8QQ, UK \\
\vspace{3mm}
{\large  C.~Morningstar} \\
University of California at San Diego  \\
La Jolla, CA 92093-0319, USA \\
\vspace{3mm}
{\large  J. Sloan~${\vphantom
{\Large A}}^3$} \\
University of Kentucky \\
Lexington, KY 40506-0055, USA \\
\end{center}
\newpage
\begin{abstract}
{We present a quenched lattice calculation of  heavy-light
meson decay constants,
using non-relativistic (NRQCD) heavy quarks in the mass region of the $b$ quark
and heavier, and clover-improved light quarks. The NRQCD Hamiltonian and the 
heavy-light current include the  corrections at first order in the 
expansion in the inverse heavy quark mass. We study the dependence of the 
decay constants on the heavy meson mass $M$, for light quarks with the tree 
level ($c_{SW}$ = 1), as well as the tadpole improved clover coefficient.
We compare decay constants from NRQCD with results from clover ($c_{SW}=1$) 
heavy quarks. 

Having calculated the current renormalisation constant $Z_A$ in one-loop
perturbation theory, we demonstrate how the heavy mass dependence of the
pseudoscalar decay constants changes after renormalisation. For the first time,
we quote a result for $f_B$ from NRQCD including the full one-loop matching factors
at $O(\alpha/M)$.}
\end{abstract}
\vspace{3cm}
\thispagestyle{empty}
\end{titlepage}
\newpage
\section{Introduction}
A calculation of the decay constant of the $B$ meson, $f_B$, is of interest 
for the determination of the unitarity triangle parameterising CP-violation in 
the Standard Model. The element $|V_{td}|$ of the Cabibbo-Kobayashi-Maskawa 
(CKM) matrix can be determined from an experimental study of
 $B-\overline{B} $ mixing, using $f_B$ as one of the input 
parameters~\cite{CKM}. It
can be defined through the following matrix element (in Minkowski space):
\beq
ip_{\mu}f_B = \langle 0|A_{\mu}|B\rangle,
\eeq
\setcounter{footnote}{0}
where $A_\mu$ is the heavy-light axial vector current. Similarly, the vector
decay constant can be defined through the relation\footnote{With this 
definition $f_V = f_B$ in the static limit}:
\beq
i\epsilon_\mu f_V M_V = \langle 0|V_\mu|B^\ast\rangle,
\eeq
$V_\mu$ being the heavy-light vector current. 
A practical tool for calculations involving hadrons containing one 
heavy quark is Heavy 
Quark Effective Theory (HQET)~\cite{HQET}. It exploits  the fact that in 
heavy-light systems, in the limit of infinite heavy quark  mass $m_Q$, there is a 
spin-flavour symmetry between heavy quarks. Corrections due to finiteness of the heavy 
quark mass
are included in an expansion in $1/m_Q$. For the decay constant one
expects a heavy mass dependence of the following kind:
\begin{equation}
f\sqrt{M} = A_0\left(1+A_1/m_Q + A_2/m_Q^2 + \ldots \right).
\end{equation}

Here, $f$ denotes the pseudoscalar or vector decay constant and $M$ the mass of the 
corresponding heavy meson.
Using lattice QCD, these matrix elements  can be calculated nonperturbatively from 
first principles.
Lattice calculations of heavy-light decay constants have been performed  using the 
relativistic `naive' (see e.g.~\cite{wupp}), or  the clover-improved 
(e.g.~\cite{kappac}), Wilson action for the heavy quarks. A reinterpretation of the 
naive relativistic action
in the regime $am_Q \gtrsim 1$ has been proposed by~\cite{FNAL}. In recent studies 
(e.g.~\cite{onogi,fB_MILC,JLQCD,APE_new}), 
this suggestion has been implemented to various degrees. These simulations have either
used heavy quarks in the mass region of the charm, relying
on extrapolations to the $b$ quark, or gone at most up to masses around the $b$. 
Lattice calculations have also been done in the 
infinite mass (static) limit (e.g. Refs.~\cite{APE_static,eichten,UKQCD_static}). For a
recent overview of the status of lattice calculations see Ref.~\cite{flynn}.

Alternatively,  heavy quarks can be simulated using 
non-relativistic QCD (NRQCD)~\cite{lepage92}, an effective theory where the 
operators in the
action  and heavy-light currents are expanded in a series in the bare inverse heavy quark
mass $\mnod$.  With this approach, one can study quarks in the whole region
between the $b$ quark and the static limit. The first calculation of
heavy-light decay constants with NRQCD~\cite{christine}, and a following 
more extensive  simulation~\cite{hashimoto} used an NRQCD action at $O(1/\mnod)$
for the heavy quark and a Wilson action for the light quark; however the
$O(1/m^{(0)}_Q)$ corrections to the current operators were not included. A
calculation using quenched configurations and Wilson light quarks that includes the
current corrections is introduced in Ref.~\cite{draper}.
The first report on 
a study of  decay constants where the currents are also corrected through
$O(1/\mnod)$, using Wilson light quarks and 
dynamical configurations, is published in~\cite{Sara}. In the study described
in this paper, we use an NRQCD action and currents as in~\cite{Sara} but with
quenched configurations and clover light quarks.

In Sec.~\ref{sec:details} we explain the operators that we use in our
NRQCD Hamiltonian and heavy-light currents, the simulation parameters and the
interpolating operators for the mesons. The fitting procedure and results for
energies and amplitudes are presented in Sec.~\ref{sec:fitting}. The 
analysis of the results  follows in Sec.~\ref{sec:results}. After a brief
discussion of the heavy-light meson masses using tree-level-improved ($c_{SW}
= 1$), and tadpole-improved light fermions, we turn to the decay matrix 
elements. We compare the decay constants with tree-level-improved light quarks
and NRQCD heavy quarks with results using a clover action with $c_{SW} = 1$  
also for the heavy quark. This is followed by a study of axial, vector,
and spin-averaged matrix elements with tree-level and tadpole-improved
 light clover quarks as a function of the heavy quark mass. We give
results for the physical ratio $f_{B_s}/f_{B_d}$. Using a renormalisation
constant $Z_A$ from one-loop perturbation theory, we finally present renormalised
axial matrix elements, and quote our estimate of $f_B$.
\section{\label{sec:details}Simulation details}
We choose to work in a Pauli basis where the two-component heavy quark
spinor $Q$ and the antiquark spinor $\tilde{Q}^\dagger$ decouple~\cite{Sara}.
Only the spinor $Q$ is used in our simulations.
The non-relativistic Lagrangian which describes
the $b$ quark is expanded through $O(1/m_Q^{(0)})$ at tree level:
\begin{equation}
{\cal L} = Q^{\dagger} \left( D_t +  H_0 + \delta H\right) Q,
\end{equation}
where
\begin{equation}
H_0 = -\frac{\Delta^{(2)}}{2m_{Q}^0},\; \delta H =
-\frac{c_B\,\vec{\sigma}\cdot\vec{ B}}{2m_Q^{(0)}}.\label{eq:terms}
\end{equation}
The gauge links are tadpole improved:
\beq
U_{\mu} \rightarrow U_{\mu}/u_0, \;\, u_0^4 = \langle 
1/3TrU_{\rm Plaq.}\rangle,
\eeq
so the covariant derivatives act in the following way on a Green 
function $G$:
\begin{eqnarray}
\Delta_\mu G(x) &=& [U_\mu(x) G(x+\hat{\mu}) - U_\mu^\dagger(x-\hat{\mu})
G(x - \hat{\mu})]/(2u_0); \\
\Delta_\mu^{(2)} G(x) &=& [U_\mu(x) G(x+\hat{\mu}) + U_\mu^\dagger(x-\hat{\mu})
G(x - \hat{\mu})]/u_0 -2G(x),
\end{eqnarray}
and
\beq
\Delta^{(2)} = \sum_\mu \Delta^{(2)}_\mu.
\eeq
We use the tree level coefficient $c_B=1$ in Eq.~(\ref{eq:terms}). Using 
tadpole-improvement, we expect that the perturbative contributions to this 
coefficient do not become too large. For $q^\ast$ values ranging between 
$1/a$ and $\pi/a$, $\alpha_V$ lies between $\sim \! 0.25$ and  
$\sim \!0.15$. The  perturbative correction to this operator may thus be 
roughly of order 20\%.
The heavy quark follows the evolution equation~\cite{upsilon}
\begin{eqnarray}
 G_1 &=&
  \left(1\!-\!\frac{aH_0}{2n}\right)^{n}
 U^\dagger_4
 \left(1\!-\!\frac{aH_0}{2n}\right)^{n} \, \delta_{\vec{x},0},\;t=1,
\end{eqnarray}
on the first timeslice, and on the following timeslices
\begin{equation}
G_{t+1} = \left(1-\frac{aH_0}{2n}\right)^n U_4^{\dagger}
\left(1-\frac{aH_0}{2n}\right)^n (1-a\delta
H)G_t ,\; t>1.
\end{equation}
It has been noted~\cite{tanmoy} that
this evolution equation introduces an error in principle of 
$O(a\Lambda^2_{QCD}/\mnod)$ in the
amplitude since the operator $\delta H$ is not applied on the first timeslice.
We estimated the actual size of the error by comparing it with a different
evolution equation and found it to be $\sim \!3-4\%$ for the bare lattice matrix 
element~\cite{saraprep} (using clover light quarks and dynamical 
configurations). This deviation is of the order of the statistical error (see 
Sec.~\ref{sec:results}), so it will be ignored here.

In the calculation of   decay constants to the desired order in the 
$1/m_Q^{(0)}$-expansion, one has to also
include the corrections to the currents. At tree level, these can be 
obtained by relating the heavy 
quark field in full QCD, $h$, and the nonrelativistic heavy  quark field 
through an inverse Foldy-Wouthuysen transformation on the heavy 
quark spinor. At $O(1/m_Q^{(0)})$, one has:
\begin{equation}
h = \left(1-iS^{(0)}\right) \left(\begin{array}{c}Q \\ \tilde{Q}^\dagger
\end{array}\right), 
\eeq
where 
\beq
S^{(0)} = -i\frac{\vec{\gamma}\cdot\vec{D}}{2m_Q^{(0)}}.
\eeq
We  write the heavy-light currents in full QCD as
\beq
A_\mu = \bar{q}\gamma_5\gamma_\mu h,\mbox{  and }V_\mu = 
\bar{q}\gamma_\mu h,
\eeq
$q$ being the light quark field. In the following we consider only the
time component of the axial vector current and the spatial components
of the vector current.
In our simulations the  conventions for the $\gamma$ 
matrices are:
\beq
\gamma_0 =\left(\begin{array}{lr}1\!\!1& 0 \\ 0 &-1\!\!1
\end{array}\right) ,\;\;\vec{\gamma} =
\left(\begin{array}{lr}0&i\vec{\sigma}
\\-i\vec{\sigma}& 0
\end{array}\right),\;\;\gamma_5 =\left(\begin{array}{lr}0&1\!\!1 \\ 
1\!\!1& 0
\end{array}\right).
\eeq
The current, corrected though $O(1/m_Q^{(0)})$, takes the form,
\begin{equation}
J_\mu = J_\mu^{(0)} + J_\mu^{(1)},
\end{equation}
where the contributions to the currents are given by:
\begin{equation}
J_5^{(0)} = q_{34}^\dagger Q, \; 
J_5^{(1)} = -i\frac{1}{2\mnod}q_{12}^\dagger\vec{\sigma}\cdot\vec{D}Q
\end{equation}
for the axial current and 
\begin{equation}
J_k^{(0)} = -i q_{34}^\dagger\sigma_k Q, \; 
J_k^{(1)} = -\frac{1}{2\mnod}q_{12}^\dagger\sigma_k
\vec{\sigma}\cdot\vec{D}Q
\end{equation}
for the vector current, where we use the notations $q_{12}$ for the upper and $q_{34}$ 
for the lower two components of the light quark spinor.
Other operators of the same mass dimension and lattice symmetry mix under 
renormalisation with the Foldy-Wouthuysen operators. For the axial current, we discuss this 
further in subsection~\ref{sec:ren}. A more general list of these operators can be found e.g. 
in Refs.~\cite{GoldenHill} and~\cite{Sara}.


We also implement 
the heavy quark in the static approximation, which corresponds to the 
Lagrangian:
\beq
{\cal L}  = Q^{\dagger} D_t Q.
\eeq
The static heavy quark propagator follows the evolution equation:
\beq
G_{t+1} - U_4^{\dagger}G_t = \delta_{x,0}.
\eeq
Our light quark propagators were generated by the UKQCD Collaboration. These
use a clover-improved Wilson formulation~\cite{SW}. In the
following we will denote our simulation with the tree level clover coefficient $c_{SW}$
= 1 as Run A and  the simulation with tadpole-improved clover fermions, $c_{SW} = 1/u_0^3$,
with Run B. For $\kappa$ values and other details of the light quarks  see 
Table~\ref{tab:details}. The light quarks in Run A are rotated~\cite{Heatlie91}:
\begin{eqnarray}
q(x) & \rightarrow &  \left(1-\frac{a}{2}\gamma\cdot D\right)q(x), \nonumber \\
\overline{q}(x) & \rightarrow &  
\overline{q}(x)\left(1+\frac{a}{2}\gamma\cdot \stackrel{\leftarrow}{D}
\right).
\end{eqnarray}
For the light quarks in Run B we use the normalisation $\sqrt{1-6\tilde{\kappa}}$.

The clover improvement removes lattice spacing errors at $O(a)$, and we expect the remaining 
leading errors for light quarks at zero momentum to be $O(\alpha_s a\Lambda_{QCD})$ and
$O(a^2 \Lambda_{QCD}^2 )$. If we use for $\Lambda_{QCD}$ a value around 300 MeV 
($a\Lambda_V$
 = 0.169 for our configurations), and $\alpha_V(1/a) = 0.247$, we estimate these
errors to be $\sim \!4\%$ and
$\sim \!3\%$ respectively.

We use quenched gauge
configurations at $\beta = 6.0$ on $16^3\times 48$ lattices, generated by the 
UKQCD Collaboration.  The configurations were fixed to Coulomb
 gauge. In Table~\ref{tab:details} we list the ensemble sizes and the lattice 
spacings from light spectroscopy. Degenerate pion and rho masses, lattice spacings and 
results for $\kappa_{\rm crit}$
and $\kappa_{\rm s}$ are taken for Run A from Ref.~\cite{kappac}, and for Run B from
Ref.~\cite{PRowland}. The heavy quark parameters for both runs are
given in the same table. The variation of the ensemble sizes in Run B  for different 
$\kappa$ and $\mnod$ values between 45 and 67 is due to limited computer time.
One has to note  that for quenched configurations ratios of physical
quantities generally differ from the corresponding ratios in the real world. 
Thus the values one obtains for the lattice spacing from different 
physical quantities are in general different. Averaging results from the 
$1P-1S$
and $2S-1S$ splitting of the $\Upsilon$, one obtains $a^{-1}$ = 2.4(1) GeV at
$\beta = 6.0$~\cite{upsilon}. Probably 
for  heavy-light systems the appropriate  lattice spacing is closer to the one
determined from light hadron spectroscopy, and in the following we will use 
$a^{-1} = 2.0(2)$ GeV to convert lattice results into physical units.
This encompasses $a^{-1}$ from $m_\rho$ for both runs (see Table~\ref{tab:details})
and also lattice spacings from $f_\pi$ (see e.g.~\cite{APE,fpi}) as well as from 
 gluonic  quantities~\cite{bali} at $\beta$ = 6.0.
The heavy quark masses 
 $am_Q^{(0)}$ = 1.71, 2.0, 2.5 are in the region of the $b$ quark. In our later
simulation (Run B) we choose one heavy mass as high as $a\mnod$ = 8.0, because
in Run A with the heaviest mass being 4.0 we find it difficult to 
extrapolate to the static limit. The value $a\mnod$ = 1.71
corresponds to the $b$ quark mass as determined from $\Upsilon$ 
spectroscopy with $a^{-1}$ = 2.4(1) GeV~\cite{upsilon,Mbpaper}.   
If one chooses the lattice spacing from light 
spectroscopy, $a^{-1} \simeq 2$ GeV, $am_Q^{(0)} = 2.0$ gives approximately  
the  same dimensionful $b$ quark mass as the value $a\mnod$ =  1.71 used in 
the $\Upsilon$ simulations at $a^{-1} = 2.4$ GeV. This ambiguity in fixing 
lattice bare quark masses is typical of problems with the quenched 
approximation.
\begin{table}
\begin{center}
\begin{tabular}{|c|c|c|}
\hline
\multicolumn{1}{|c}{} &
\multicolumn{1}{|c}{Run A} &
\multicolumn{1}{|c|}{Run B} \\
\hline
\multicolumn{1}{|c}{} &
\multicolumn{2}{|c|}{Light quarks} \\
\hline
 $c_{SW}$& 1.0  & $1/u_0^3$ \\
rotated &  yes & no \\
$\begin{array}{c}\kappa \mbox{ values}\\ am_\pi \\ am_\rho\end{array}$& 
$\begin{array}{c|c}0.1432& 0.1440\\0.386({\raisebox{0.08em}
{\scriptsize {$\;\begin{array}{@{}l@{}}
\makebox[15pt][c]{$+$}\makebox[0.9em][r]{4} \\[-0.12em]
\makebox[15pt][c]{$-$}\makebox[0.9em][r]{4} \end{array}$}}})&0.311({\raisebox{0.08em}
{\scriptsize {$\;\begin{array}{@{}l@{}}
\makebox[15pt][c]{$+$}\makebox[0.9em][r]{6} \\[-0.12em]
\makebox[15pt][c]{$-$}\makebox[0.9em][r]{5} \end{array}$}}})\\
0.51({\raisebox{0.08em}
{\scriptsize {$\;\begin{array}{@{}l@{}}
\makebox[15pt][c]{$+$}\makebox[0.9em][r]{2} \\[-0.12em]
\makebox[15pt][c]{$-$}\makebox[0.9em][r]{1} \end{array}$}}})&0.47({\raisebox{0.08em}
{\scriptsize {$\;\begin{array}{@{}l@{}}
\makebox[15pt][c]{$+$}\makebox[0.9em][r]{3} \\[-0.12em]
\makebox[15pt][c]{$-$}\makebox[0.9em][r]{2} \end{array}$}}})\end{array}$ & 
$\begin{array}{c|c} 0.1370& 0.1381 \\0.4137({\raisebox{0.08em}{\scriptsize
{$\;\begin{array}{@{}l@{}}
\makebox[15pt][c]{$+$}\makebox[0.9em][r]{11} \\[-0.12em]
\makebox[15pt][c]{$-$}\makebox[0.9em][r]{9} \end{array}$}}})& 0.2940({\raisebox{0.08em}
{\scriptsize {$\;\begin{array}{@{}l@{}}
\makebox[15pt][c]{$+$}\makebox[0.9em][r]{13} \\[-0.12em]
\makebox[15pt][c]{$-$}\makebox[0.9em][r]{12} \end{array}$}}})\\
0.538({\raisebox{0.08em}{\scriptsize
{$\;\begin{array}{@{}l@{}}
\makebox[15pt][c]{$+$}\makebox[0.9em][r]{3} \\[-0.12em]
\makebox[15pt][c]{$-$}\makebox[0.9em][r]{2} \end{array}$}}}) & 0.463({\raisebox{0.08em}
{\scriptsize
{$\;\begin{array}{@{}l@{}}
\makebox[15pt][c]{$+$}\makebox[0.9em][r]{6} \\[-0.12em]
\makebox[15pt][c]{$-$}\makebox[0.9em][r]{4} \end{array}$}}})\end{array}$\\
$\kappa_{\rm crit}$ & 0.14556(6) &0.13926(1) \\
$\kappa_{\rm s}$ & 0.1437({\raisebox{0.08em}{\scriptsize
{$\;\begin{array}{@{}l@{}}
\makebox[15pt][c]{$+$}\makebox[0.9em][r]{4} \\[-0.12em]
\makebox[15pt][c]{$-$}\makebox[0.9em][r]{5} \end{array}$}}})($m_K$) & 0.13758(5)($m_K$), 
0.13726(9)($m_\phi$) \\
$a^{-1}(m_\rho)$& 2.0(2) GeV~\cite{kappac} &1.99(4) GeV~\cite{PRowland} \\
\hline
\multicolumn{1}{|c}{} &
\multicolumn{2}{|c|}{Heavy quarks} \\
\hline
$\begin{array}{c}a\mnod\\n\end{array}$ & 
$\begin{array}{c|c|c|c}1.71 &2.0 &2.5 & 4.0\\2&2&2&2\end{array}$&
$\begin{array}{c|c|c|c} 1.71& 2.0& 4.0& 8.0 \\2&2&1&1\end{array}$\\
\hline
\multicolumn{1}{|c}{} &
\multicolumn{2}{|c|}{Configurations} \\
\hline
number & 35 + time reversed & 45 -- 67 \\
\hline
\end{tabular}
\end{center}
\caption{Simulation details.}
\label{tab:details}
\end{table}

At the source we use the following interpolating fields for the mesons:
\beq
\sum_{\vec{x}_1,\vec{x}_2}Q^{\dagger}(\vec{x}_1)\Gamma^\dagger(\vec{x}_1
-\vec{x}_2) q(\vec{x}_2),
\eeq
where $\Gamma(\vec{x}_1-\vec{x}_2)$ factorizes into a smearing
function $\phi(r=|\vec{x}_1 - \vec{x}_2|)$ and one of the
$2\times 4$ matrices in spinor space shown in table~\ref{table:states}.
\begin{table}
\begin{center}
\begin{tabular}{|c|c|}
\hline
$^1S_0$ & $^3S_1$ \\
\hline
$\left(\begin{array}{c}0\\1\!\!1\end{array}\right)$ &
 $\left(\begin{array}{c}0  \\ \sigma\end{array}\right)$ \\
\hline
\end{tabular}
\end{center}
\caption{Spin operators for mesonic states.}
\label{table:states}
\end{table}
We calculate heavy-light current matrix elements with the smearing function 
$\phi$ being either a delta function or a ground or excited state 
hydrogen-like wave function~\cite{charmonium} at  the source and a delta 
function at the sink, and
meson correlators with all combinations of smearing functions at source and
sink. For the ground state, we use 
\beq
\phi(r) = \exp(-r/r_0),
\eeq
and for the radially excited state,
\beq
\phi(r) = \left(\frac{1}{2}\right)^{3/2}\left(1-\frac{r}{2r_0}\right)
\exp(-r/(2r_0)),
\eeq
choosing $ar_0 = 3$. The smearing is applied on the heavy quark. 
\section{\label{sec:fitting}Fitting procedure}
In this section, the lattice spacing $a$ is set to 1.
On the lattice, the decay constant can be extracted from 
the matrix element  of the local  current $J_k$:
\begin{equation}
f M = \langle 0|J_k|B \rangle,
\end{equation}
with $k = 5$ for pseudoscalar  and $k=1,2,3$  for vector decay constants. $M$  
denotes the meson mass.
The correlation function of this current decays for sufficiently large times
exponentially:
\begin{equation}
C_{LL}(t) \rightarrow Z_L^2 e^{-E_{\rm sim} t}
\end{equation}
We are using the notation $C_{rs}$ for the correlation 
functions, where the index $r$ denotes the smearing function at the 
source, and the index $s$ the  smearing function at the sink.
$L$ stands for a delta function, 1 for the ground state and 2 for the 
excited state smearing function.
$E_{\rm sim}$ is the bare ground state binding energy and $Z_L$ is related to
the matrix element as follows:
\begin{equation}
Z_L = \frac{1}{\sqrt{2M}}\langle 0|J_k|B \rangle,
\end{equation}
so that
\begin{equation}
f\sqrt{M} = \sqrt{2} Z_L.
\end{equation}
To distinguish between the uncorrected current and the current containing
the first order correction in the 
$1/\mnod$-expansion, we will use the definitions
\beq
(f\sqrt{M})^{uncorr} \equiv \frac{1}{\sqrt{M}} \langle 0|J_k^{(0)}|B\rangle,
\eeq
and
\beq
f\sqrt{M} \equiv \frac{1}{\sqrt{M}} \langle 0|(J_k^{(0)}
+J_k^{(1)})|B\rangle.
\eeq
The current corrections we denote as 
\beq
\delta(f\sqrt{M}) \equiv \frac{1}{\sqrt{M}} \langle 0|(J_k^{(1)}|B\rangle.
\eeq
Local meson operators, however, overlap considerably with excited
states with the same quantum numbers. We therefore
smear the heavy quark field at the source. The amplitude of the corresponding
correlation function then contains also the density $Z_S$, which we
determine separately from smeared-smeared correlators. 
 For sufficiently large times one has:
\begin{eqnarray}
C_{11} & \rightarrow &  A_{11} e^{-E_{\rm sim}t},\; A_{11} = Z_S^2,\label{eq:SS} \\
C_{1L} & \rightarrow &  A_{1L} e^{-E_{\rm sim}t},\; A_{1L} = Z_S Z_L.\label{eq:SL}
\end{eqnarray}
We extract the
matrix element $Z_L$ by fitting $C_{1L}$ and $C_{11}$ simultaneously
to a single exponential, using a bootstrap procedure as described 
in~\cite{us_quenched}. $Z_L$
is calculated as the correlated ratio $A_{1L}/\sqrt{A_{11}}$.
\subsection{Run A}
For $t_{min} < 5$, no single exponential fit to both 
$C_{1L}$ and $C_{11}$ is possible. After $t_{min} = 5$ the amplitudes $A_{11}$ 
and $A_{1L}$ and the ground state energy still decrease slightly, until 
$t_{min}$ is moved out to 9. 
The dependence of results on the fit interval is shown for 
$\mnod$ = 4.0 in Table~\ref{table:results_var_tmin}. For 
 $\kappa$ = 0.1440  the decrease may be by more than $1\sigma$. 
We use a fit interval $t_{min}/t_{max} = 9/25$ for all $\kappa$ and
heavy mass values. Given that the statistics are poor, we have to discard the 
smallest eigenvalues of the covariance matrix in the singular value decomposition
(SVD) algorithm. The fit results for the amplitudes from Run A are shown in 
Table~\ref{table:fit_ampl}.  
\begin{table}
\begin{center}
\begin{tabular}{|r|c|c|c|c|c|c|}
\hline
                   \multicolumn{1}{|c|}{}
                 & \multicolumn{3}{c|}{$\kappa=0.1432$}
                 & \multicolumn{3}{c|}{$\kappa=0.1440$} \\
\hline
 \multicolumn{1}{|r|}{$t_{min}/t_{max}$}
 & \multicolumn{1}{c|}{$E_{\rm sim}$}
 & \multicolumn{1}{c|}{$A_{11}$}
 & \multicolumn{1}{c|}{$A_{1L}$}
 & \multicolumn{1}{c|}{$E_{\rm sim}$}
 & \multicolumn{1}{c|}{$A_{11}$}
 & \multicolumn{1}{c|}{$A_{1L}$} \\
\hline
 $5/25$  &0.516(6)& 211(12)  & 2.37(10) &0.497(6)& $202(11)$& 2.15(11)  \\
 $6/25$  &0.515(5)& 210(11)  & 2.37(9)  &0.495(7)& $197(13)$& 2.11(13)   \\
 $7/25$  &0.514(6)& 210(13)  & 2.34(14) &0.493(8)& $194(12)$& 2.05(14)    \\
 $8/25$  &0.514(7)& 209(15)  & 2.30(17) &0.491(7)& $191(13)$& 2.00(15)    \\
  $9/25$ &0.513(7)& 210(15)  & 2.28(19) &0.491(8)& $189(13)$& 1.97(16)    \\
 $10/25$ &0.511(5)& 209(14)  & 2.26(18) &0.489(7)& $185(14)$& 1.91(17)     \\
\hline
\end{tabular}
\end{center}
\caption{Dependence of fit results from Run A  at $m_Q^{(0)}=4.0$ on the fit
interval. All quantities are in lattice units.
}
\label{table:results_var_tmin}
\end{table}
With our limited statistics, our bootstrap procedure generates
certain  ensembles on which  multi-exponential fits to two correlators fail, 
but with the original ensemble of correlation functions we
can do a simultaneous fit of $C_{1L}$ and $C_{2L}$ to 2 exponentials.
This gives a value for the ground state energy which tends to be a
little higher, but in general still compatible within one
standard deviation with the results fitted with just one 
exponential~\cite{us_quenched}.
\begin{figure}[pthb]         
\vspace{-2cm}
\centerline{\epsfxsize=8cm
\hspace{0.6cm}\epsfbox{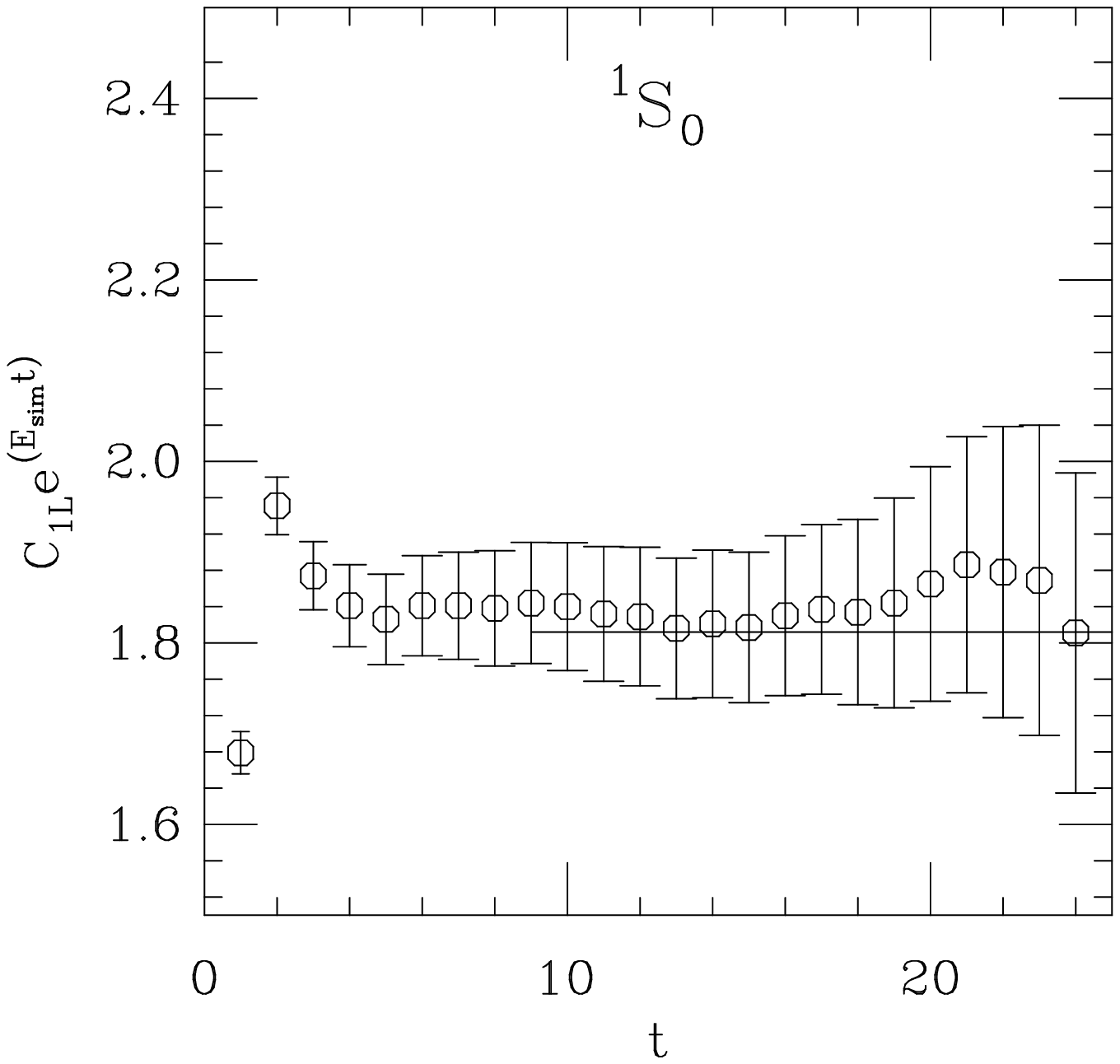}
\epsfxsize=8cm
\hspace{-0.4cm}\epsfbox{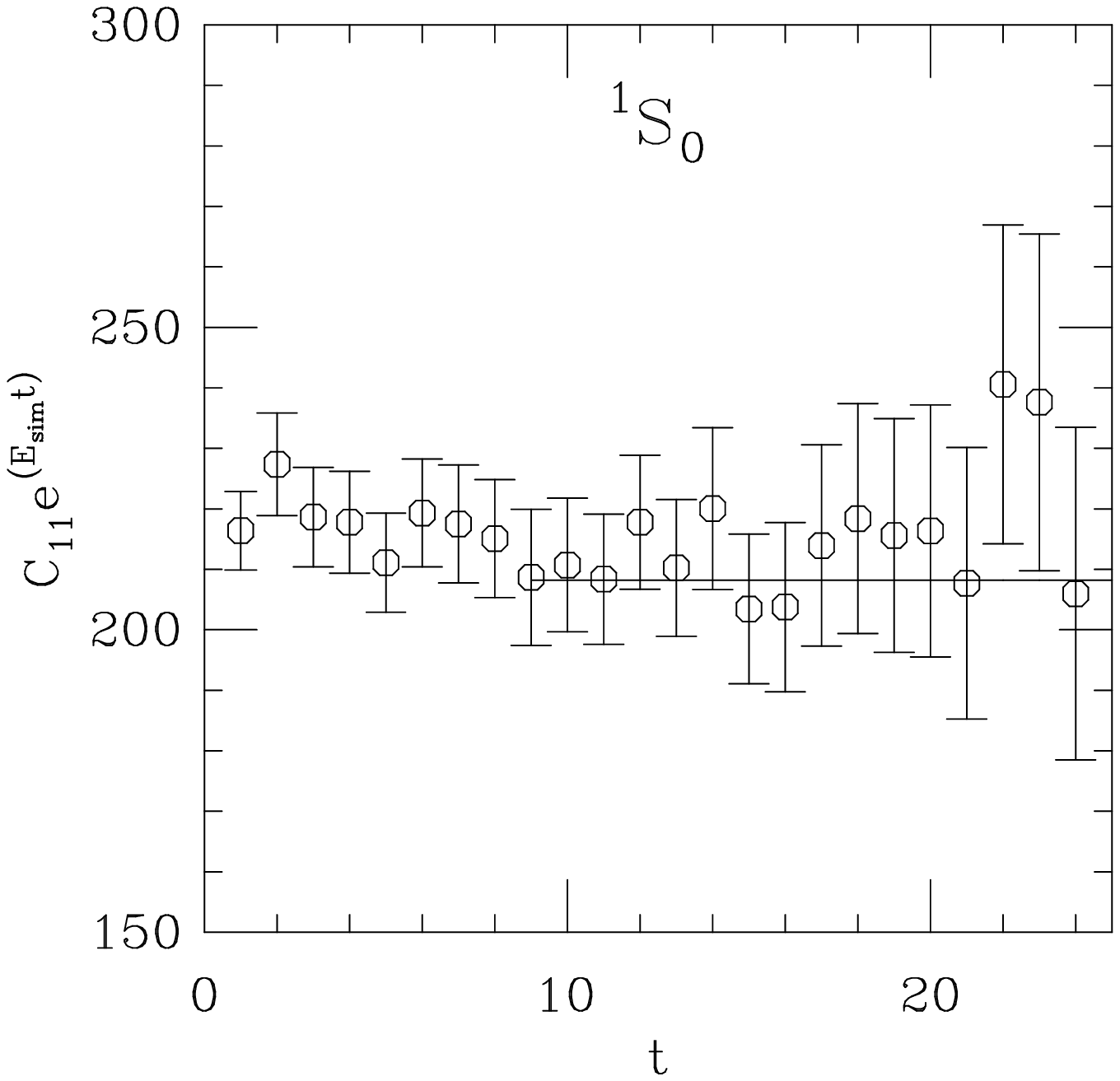}}
\vspace{-1cm}
\centerline{\epsfxsize=8cm
\hspace{0.6cm}\epsfbox{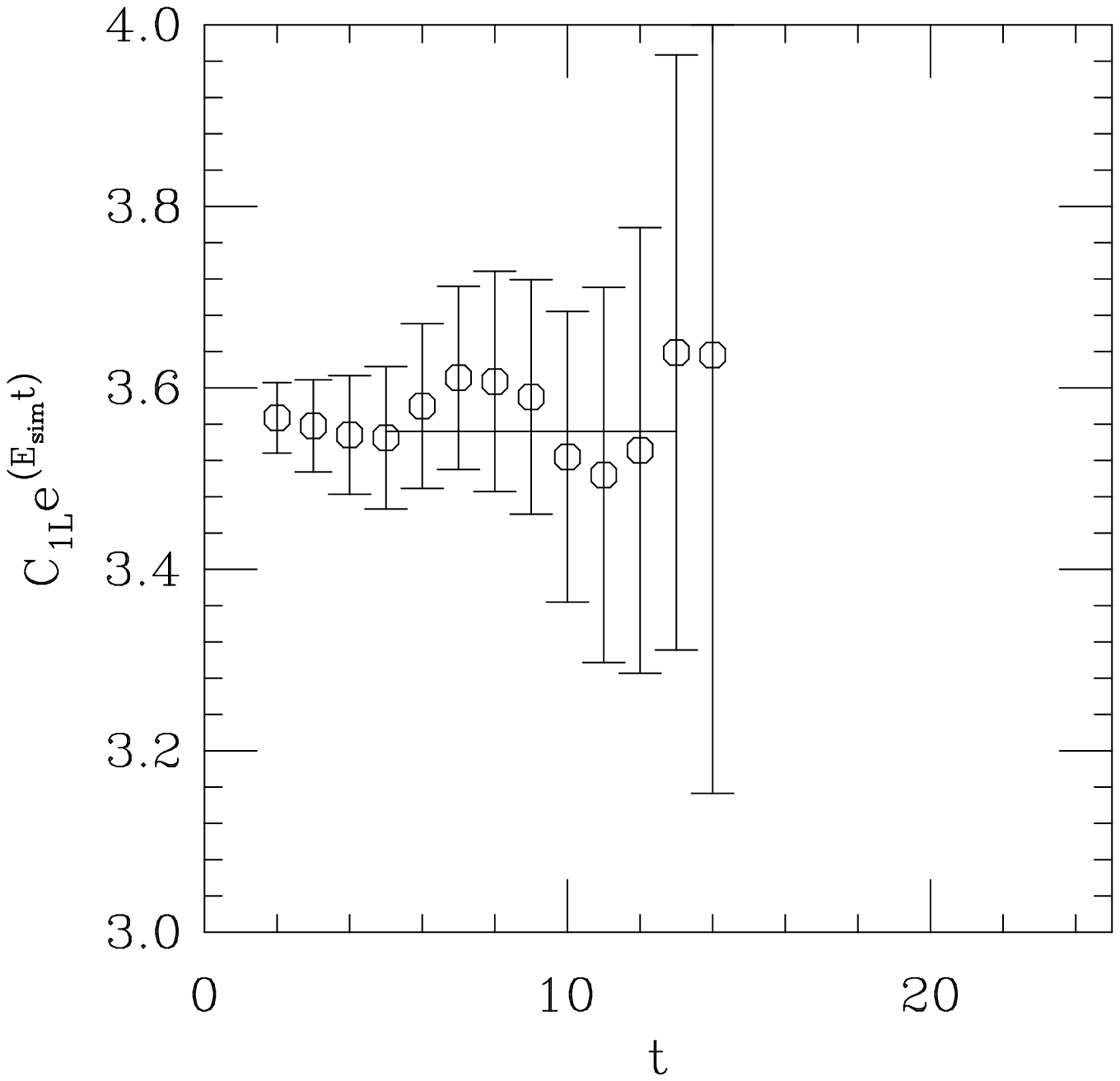}
\epsfxsize=8cm
\hspace{-0.4cm}\epsfbox{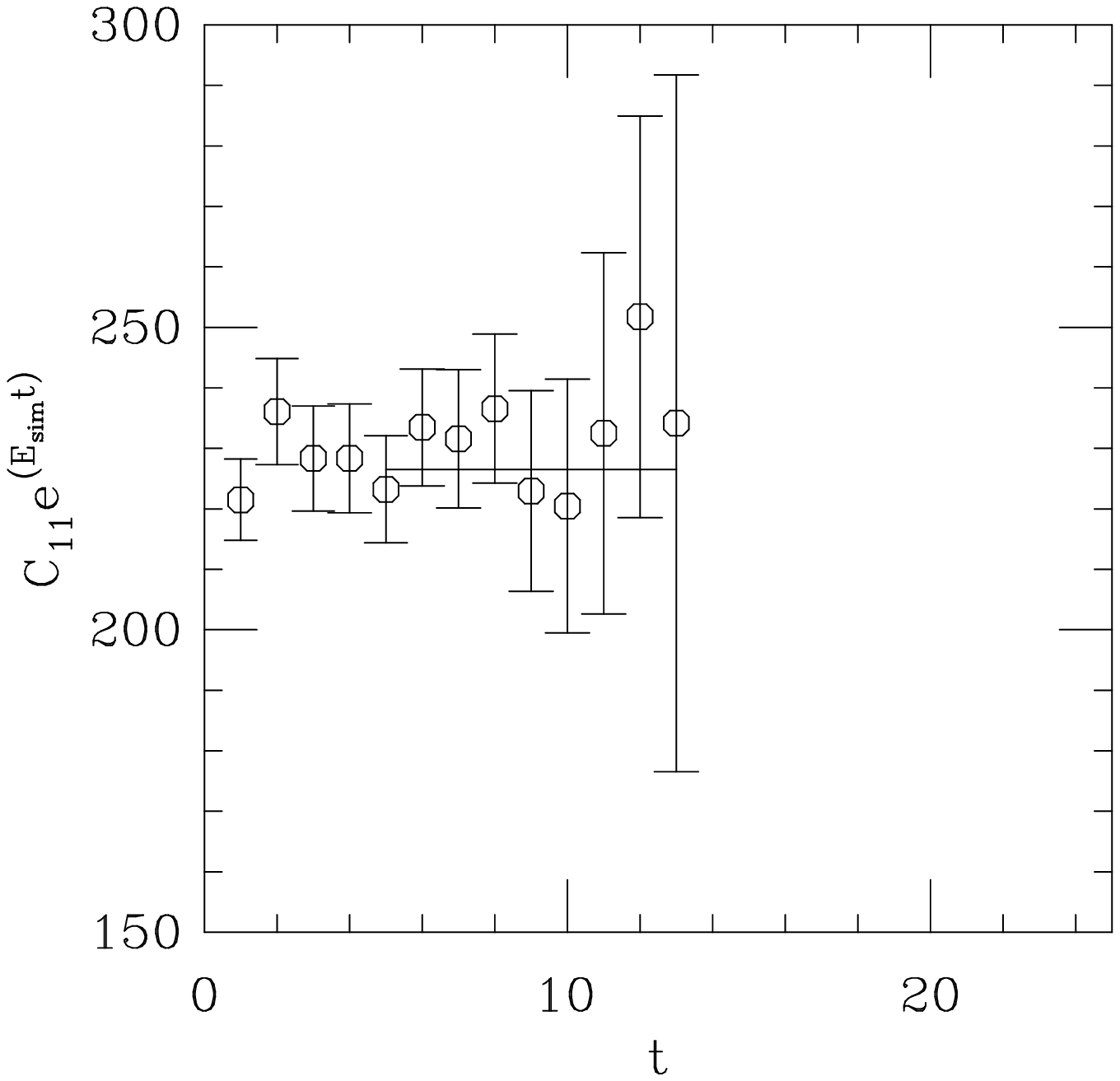}}
  \caption{Effective amplitudes for smeared-local (left) and smeared-smeared
correlators (right) from Run A;  $\kappa = 0.1432$.
The upper row shows  the $^1S_0$ case at $\mnod = 2.0$ with the $1/m^{(0)}_Q$ 
correction to the current
included, and the lower row, the static case. We choose the same scale of the x axis for
the static correlators, to demonstrate the difference in the length of the plateau between 
NRQCD and static heavy quarks.}
 \label{fig:ampl_static}
\end{figure}
\begin{table}
\begin{center}
\begin{tabular}{|c|c|l|l|l|l|}
\hline
                   \multicolumn{2}{|c|}{}
	          &\multicolumn{2}{c|}{$\kappa=0.1432$}
                  &\multicolumn{2}{c|}{$\kappa=0.1440$} \\
\hline
                   \multicolumn{1}{|c|}{$C_{1L}$}
                 & \multicolumn{1}{l|}{$m_Q^{(0)}$} 
                 & \multicolumn{1}{l|}{$A_{11}$} 
                 & \multicolumn{1}{l|}{$A_{1L}$} 
                 & \multicolumn{1}{l|}{$A_{11}$} 
                 & \multicolumn{1}{l|}{$A_{1L}$} 
\\
\hline                 
$J_5^{(0)}$ & 1.71  & 207(8) &  2.13(11)   & 197(8)   & 1.91(11)   \\
            & 2.0   & 206(8) &  2.17(10)   & 195(14)  & 1.96(15)   \\
            & 2.5   & 212(7)  & 2.33(10)    & 193(15)  & 2.03(17)  \\
            & 4.0   & 216(12) & 2.59(17)   & 199(14)  & 2.27(19)   \\
            & static& 222(6)  &   3.49(11) & 219(10)  &   3.29(14)  \\
\hline
$J_5^{(0)}+J_5^{(1)}$& 1.71& 202(16) &1.66(9) & 188(18)& 1.48(19)  \\  
                                     & 2.0 & 208(13) &1.81(10)  & 189(15)& 1.58(11)   \\
                                     & 2.5 & 210(11) &1.99(12)& 188(15) & 1.71(13)  \\
                                     & 4.0 & 210(15) &2.28(19)& 189(13) & 1.97(16) \\ 
\hline
$J_{k}^{(0)}$& 1.71  & 215(20) &   2.02(18)   & 208(16) & 1.83(15)  \\
               & 2.0   & 216(7)  &   2.10(9)    & 201(9)  & 1.87(10)  \\
               & 2.5   & 218(9)  &   2.24(13)   & 202(5)  & 1.98(10)  \\
               & 4.0   & 226(24) &   2.60(10)   & 209(8)  & 2.29(15)  \\
\hline
\end{tabular}					       
\end{center}					       
\caption{Fit results for $A_{11}$ and $A_{1L}$ from Run A.
The columns on the left indicate the operators included in the
currents.}
\label{table:fit_ampl}
\end{table}
The plateau of the static correlation functions sets in around $t_{min} =3$
and persists for about 10 timeslices, which indicates that our ground state
smearing functions work well for the static case. 

We determine the ratio of the axial current correction to the uncorrected axial
current from a fit of the bootstrap ratio of the correlators of $\langle 0
|J_5^{(0)}|B \rangle$ and $\langle 0|J_5^{(1)}|B \rangle $ to a 
constant. As fitting interval we choose $t_{min}/t_{max}=9/25$.
Results are given in Table~\ref{table:ratios_B}.
\subsection{Run B\label{sec:runb}}
The correlators from Run B are slightly noisier, as shown in the examples
in Fig.~\ref{fig:ampl_tadpole}. Also here we discard the lowest eigenvalues
of the covariance matrix in our SVD inversions. For $\kappa$ = 
0.1370, we choose $t_{min}$ = 6, for $\kappa$ = 0.1381, $t_{min}$ = 5. The
upper end of the fitting window is, for $\mnod$ = 1.71, 2.0, and 4.0,
 set to $t_{max}$ = 20, since for larger
times the signal disappears. For $\mnod$ = 8.0, the signal disappears earlier,
so we choose for $\kappa = 0.1370$, $t_{max}$ = 18 and for $\kappa$ = 0.1381, 
$t_{max}$ = 15. The goodness-of-fit 
value $Q$  of the fits is generally low ($Q \leq 0.1$). The fit results for the 
amplitudes with and without the current corrections 
are shown in Table~\ref{table:ampl_tadpole_new}. 

\begin{figure}[pthb] 
\vspace{-2cm}
\centerline{\epsfxsize=8cm
\hspace{0.6cm}\epsfbox{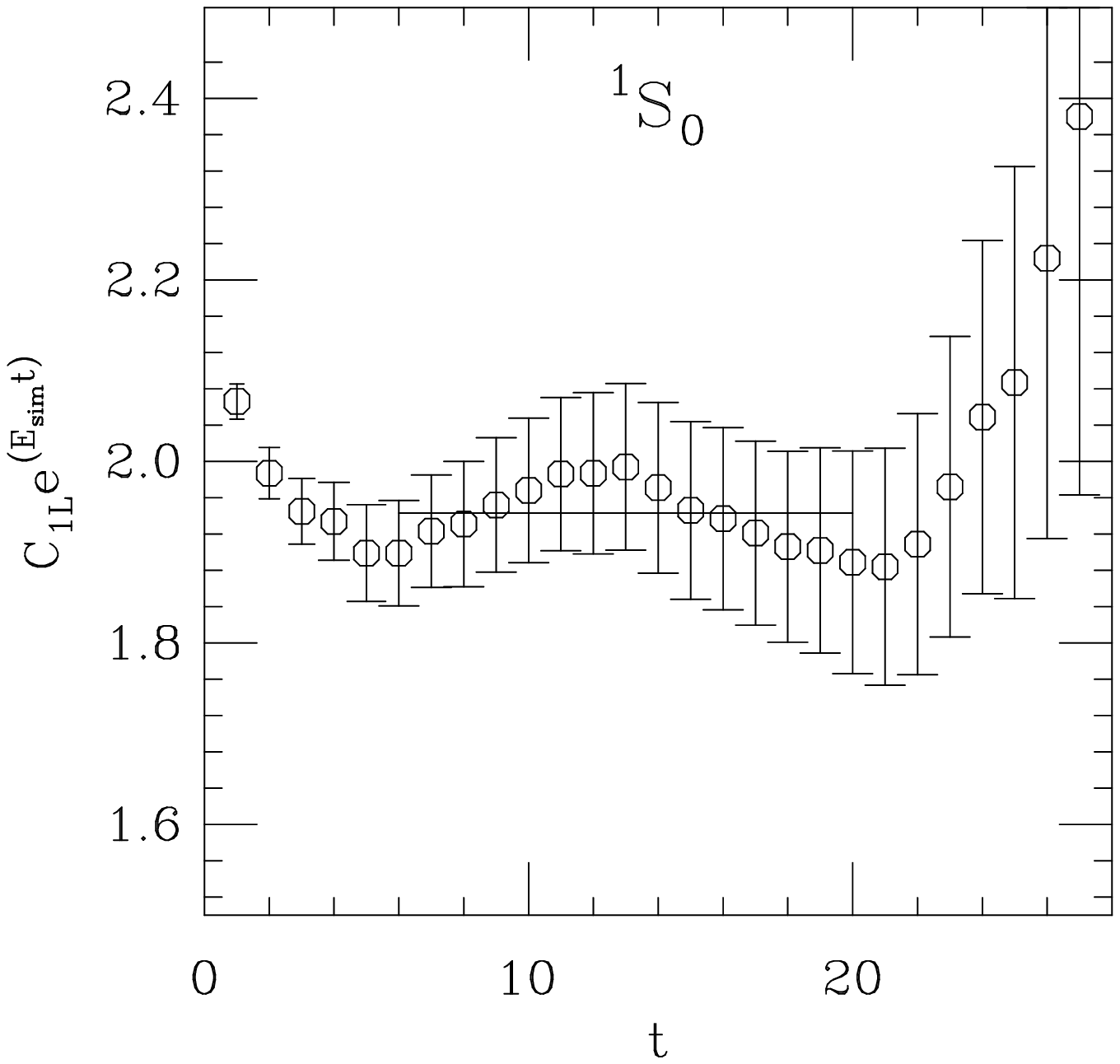}
\epsfxsize=8cm
\hspace{-0.4cm}\epsfbox{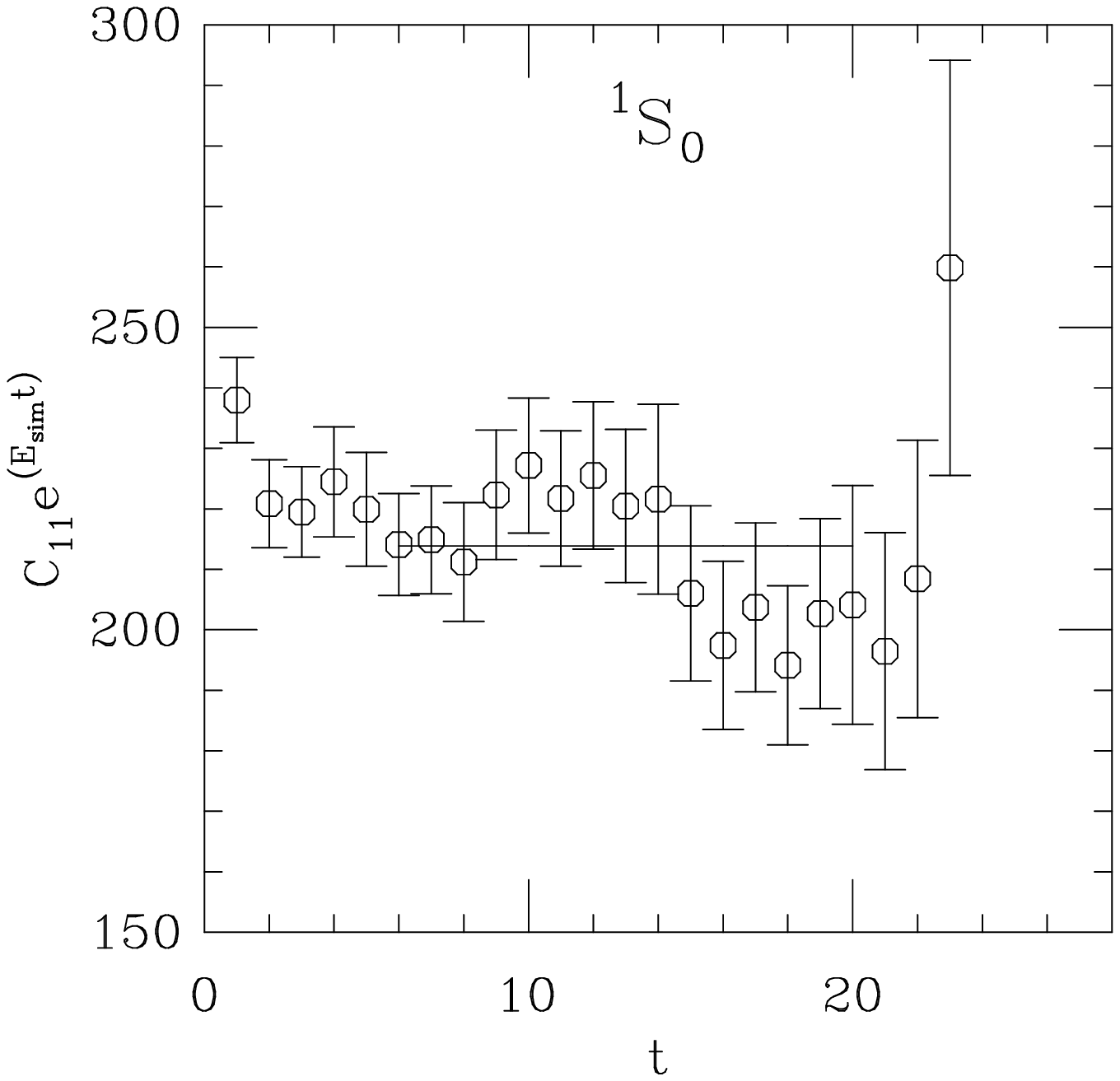}}
\vspace{-1cm}
\centerline{\epsfxsize=8cm
\hspace{0.6cm}\epsfbox{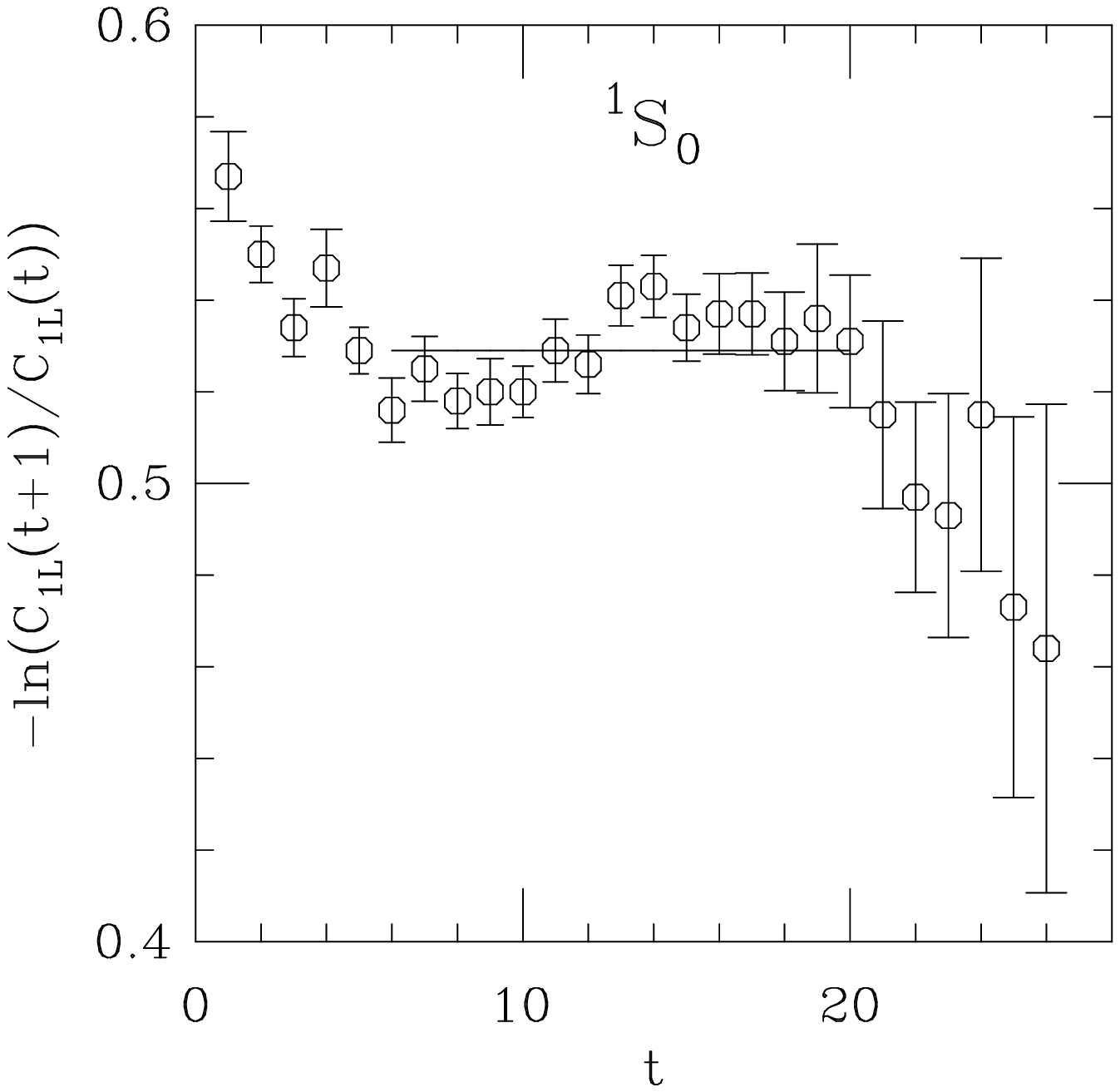}
\epsfxsize=8cm
\hspace{-0.4cm}\epsfbox{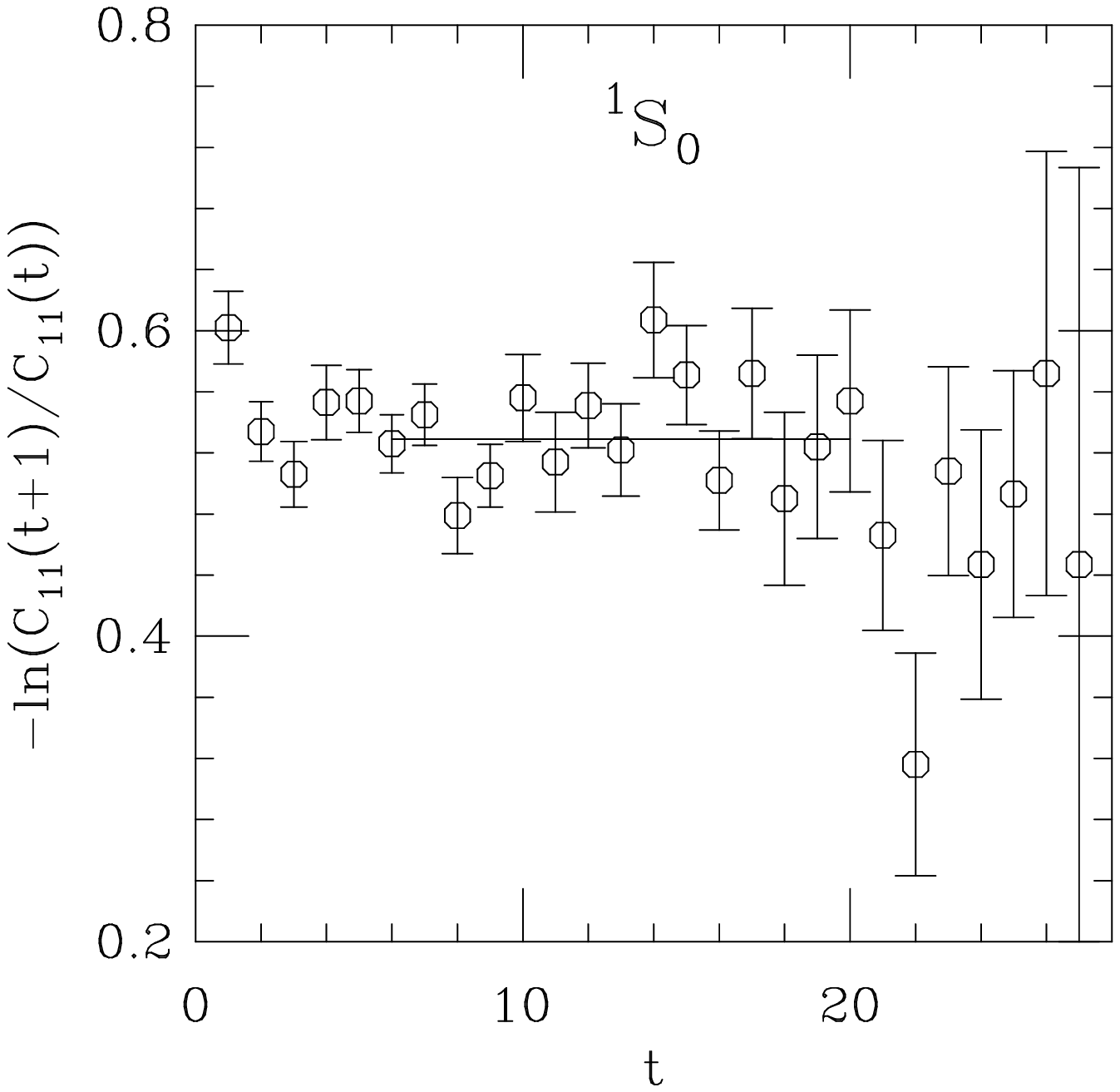}}
  \caption{NRQCD correlators from Run B at $\mnod = 2.0$ and $\kappa = 0.1370$
with  smeared source and local sink
(left),  and smeared source and sink (right). The upper row shows effective
amplitudes, the lower row, effective masses. The smeared-local correlators 
include  the $1/m^{(0)}_Q$ correction to the current. }
 \label{fig:ampl_tadpole}
\end{figure}
\begin{table}
\setlength{\tabcolsep}{0.3pc}
\begin{center}
\begin{tabular}{|c|c|l|l|l|c|l|l|}
\hline
                   \multicolumn{2}{|c|}{}
	          &\multicolumn{3}{c|}{$\kappa=0.1370$}
                  &\multicolumn{3}{c|}{$\kappa=0.1381$} \\
\hline
                   \multicolumn{1}{|c|}{$C_{1L}$}
                 & \multicolumn{1}{l|}{$m_Q^{(0)}$} 
                 & \multicolumn{1}{l|}{$E_{\rm sim}$} 
                 & \multicolumn{1}{l|}{$A_{11}$} 
                 & \multicolumn{1}{l|}{$A_{1L}$} 
                 & \multicolumn{1}{l|}{$E_{\rm sim}$} 
                 & \multicolumn{1}{l|}{$A_{11}$} 
                 & \multicolumn{1}{l|}{$A_{1L}$} \\ 
\hline                 
$J_5^{(0)}$ & 1.71  &0.524(4) & 205(7) & 2.06(7) &0.503(5) & 209(7) & 1.99(9)  \\
            & 2.0   &0.529(4) & 205(6) & 2.12(7) &0.503(6) & 208(8) & 2.02(9)   \\
            & 4.0   &0.540(5) & 203(7) & 2.32(10)&0.516(6) & 213(5) & 2.25(8)   \\
            & 8.0   &0.540(6) & 204(9) & 2.54(13)&0.511(8) & 204(7) &2.39(11)   \\
            & static&  0.540(6) &220(5) & 2.98(8) & 0.517(7)&213(9) & 2.76(9)   \\
\hline 
 $J_5^{(0)}+J_5^{(1)}$  & 1.71&0.525(4) &212(7) &1.84(6) &0.503(5)  &210(6) &1.72(7) \\
 & 2.0 &0.529(4) &214(7) &1.94(6) &0.503(6)  &208(8) &1.76(7)  \\
 & 4.0 &0.541(5) &218(5) &2.23(8) &0.516(6)  &213(6) &2.11(8)  \\
 & 8.0 &0.542(8) &218(8) &2.62(8) &0.513(8)  &211(8) &2.37(11) \\ 
\hline  				      
$J_{k}^{(0)}$& 1.71 &0.546(5) &218(8) &2.02(8)  &0.526(7) &215(3)  &1.95(9) \\
               & 2.0  &0.548(5) &218(6) &2.10(7)  &0.519(7) &214(5)  &1.93(8) \\
               & 4.0  &0.552(6) &217(7) &2.35(10) &0.524(7) &215(5)  &2.19(8) \\
               & 8.0  &0.550(6) &219(6) &2.62(10) &0.518(10) &216(3)  &2.49(9) \\
\hline 
$J_{k}^{(0)}+J_k^{(1)}$ & 1.71&0.544(4) &220(5) &2.14(6) & 0.522(6)&216(7)&2.00(8) \\ 
& 2.0 &0.546(4) &220(3) &2.21(5) & 0.519(6)&214(7)&2.02(8) \\ 
                                           & 4.0 &0.550(5) &220(6) &2.46(9) & 0.528(7)&216(4)&2.29(8) \\ 
                                           & 8.0 &0.547(6) &221(5) &2.67(8) & 0.518(8)&216(3)&2.51(9) \\ 
\hline			       			       
\end{tabular}					       
\end{center}					       
\caption{Fit results for binding energies, $A_{11}$ and $A_{1L}$  from Run B.  
The column on the left indicates the operators contributing to the currents.
All quantities are in lattice units.}
\label{table:ampl_tadpole_new}
\end{table}
At $\mnod = 1.71$, 2.0, and 4.0, the $C_{1L}$ correlation functions
from Run B have  after $t \simeq 14$ ($\kappa = 0.1370$) or
$t \simeq 12$ ($\kappa = 0.1381$) a wiggle of $\sim 1\sigma$. If we choose $t_{max}$ for 
the fits not to include this 
wiggle, we obtain a higher $Q$ ($\sim 0.3 - 0.4$), but the  results for $E_{\rm sim}$ and 
$A_{1L}$ 
are up to $\sim \!2\sigma$ smaller than the values listed in 
Table~\ref{table:ampl_tadpole_new}. We choose to extract the decay constant using the
larger fitting range. However, this variation of the result with
the fit range indicates that there is a fitting uncertainty  in $E_{\rm sim}$ and 
$A_{1L}$, associated with the choice of the fitting range, which could be for 
$\mnod = 1.71$, 
2.0 and 4.0 about twice as large  as the bootstrap errors. This propagates
into a fitting uncertainty of $\leq 2\sigma$ for the pseudoscalar
matrix elements, and of $\sim \!2-3\sigma$ for the vector matrix element
(for chirally extrapolated light quarks it is in both cases $\leq 1\sigma$).
In the tables and figures, we give pure bootstrap errors on the results from 
Run B. We will include the fitting uncertainty where we quote 
our final results for decay constants (subsection~\ref{sec:ren}). Note that the fitting
uncertainty for the static case and for $\mnod$ = 8.0 cannot be estimated in a
similar way since the signal from the correlator disappears much earlier. 

In a similar way to Run A, the plateau in the 
static correlation functions from Run B sets in  slightly earlier and is clearly
shorter than for the NRQCD correlators. For $\kappa$ = 0.1370 we use the
fit interval $t_{min}/t_{max}$ = $3/10$ and for $\kappa$ = 0.1381,
$t_{min}/t_{max}$ = $3/9$. Effective amplitude plots are shown 
in Fig.~\ref{fig:ampl_tadpole_static}.

We also determine the ratio of the current corrections to the uncorrected current from the
bootstrap ratio of their smeared-local correlators. For the ratios at $\kappa$ = 0.1370 
we choose the same fit intervals as described in 
the previous paragraph for the  correlation functions. The results are shown in 
Table~\ref{table:ratios_B}. The ratios at $\kappa$ = 0.1381 plateau later than the
correlation functions, thus we use $t_{min}=6$ instead of 5.
\begin{figure}[pthb]         
\vspace{-1.5cm}
\centerline{\epsfxsize=8cm
\hspace{0.6cm}\epsfbox{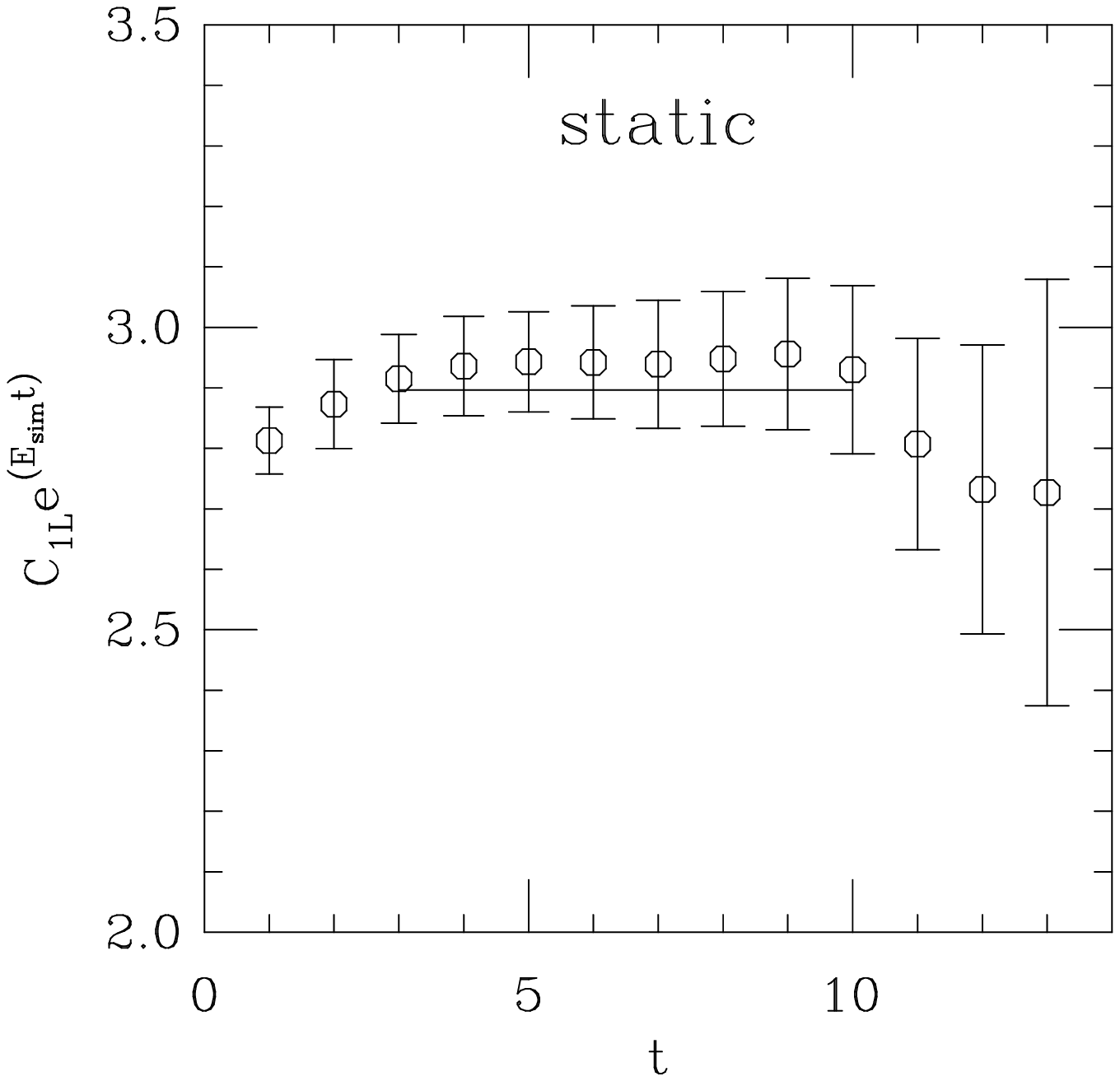}
\epsfxsize=8cm
\hspace{-0.4cm}\epsfbox{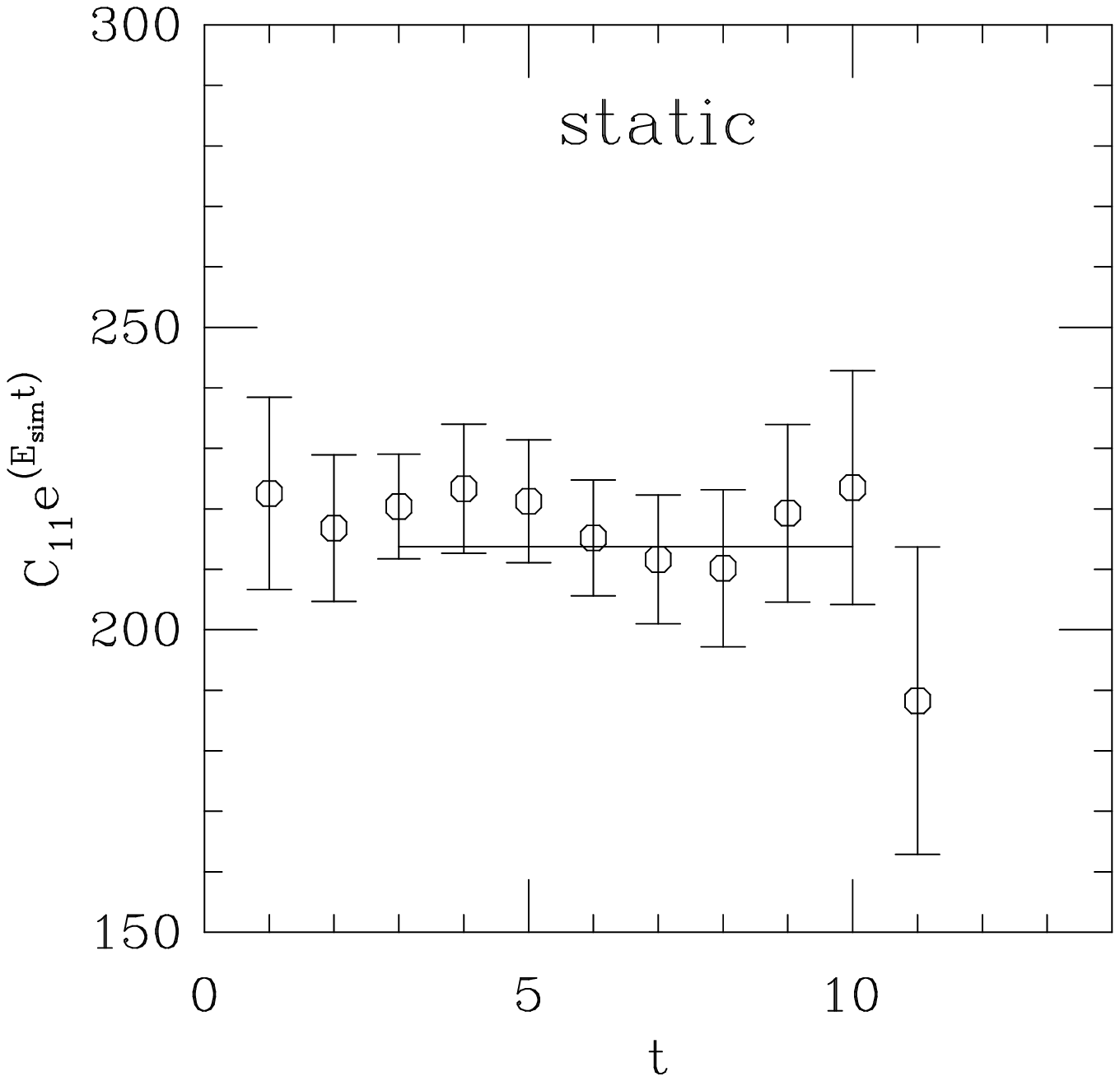}}
\vspace{-0.5cm}
\caption{Static effective amplitudes from Run B at $\kappa = 0.1370$.}
 \label{fig:ampl_tadpole_static}
\end{figure}
\begin{table}
\begin{center}
\begin{tabular}{|l|l|l|l|l|}
\hline
\multicolumn{1}{|c}{} &
\multicolumn{2}{|c|}{Run A} &
\multicolumn{2}{|c|}{Run B} \\
\hline
\multicolumn{1}{|c}{} &
\multicolumn{1}{|c|}{$\kappa$ = 0.1432} &
\multicolumn{1}{c|}{$\kappa$ = 0.1440} &
\multicolumn{1}{c|}{$\kappa$ = 0.1370} &
\multicolumn{1}{c|}{$\kappa$ = 0.1381} \\
\hline
\multicolumn{1}{|c}{$\mnod$} &
\multicolumn{4}{|c|}{$|\delta(f\sqrt{M})_{PS}/(f\sqrt{M})_{PS}^{uncorr}
|$} \\
\hline
 1.71 & 0.1964(6)   & 0.1986(5) & 0.1372(7) & 0.1438(7) \\
 2.0  & 0.1696(2)   & 0.1713(4) & 0.1215(5) & 0.1239(6) \\
 2.5  & 0.1379(3)   & 0.1389(4) &           &           \\ 
 4.0  & 0.08800(13)&0.08863(4) &0.06561(18)&0.0672(3) \\
 8.0  &            &          &0.003368(8)&0.03447(11) \\
\hline
\multicolumn{1}{|c}{$\mnod$} &
\multicolumn{4}{|c|}{$|\delta(f\sqrt{M})_{V}/(f\sqrt{M})_{V}^{uncorr}
|$} \\
\hline
1.71 & & & 0.0455(2)   & 0.0468(3) \\
2.0  & & & 0.04032(17) & 0.0410(3) \\
4.0  & & & 0.02178(5)  & 0.0223(7) \\
8.0  & & & 0.01118(3)  & 0.01151(3) \\
\hline
\end{tabular}
\end{center}
\caption{Fit results for the ratio of the current correction to the 
uncorrected matrix element, in lattice units. $\delta(f\protect\sqrt{M})_V$ 
has not been calculated  directly in Run A.}
\label{table:ratios_B}
\end{table}
\section{\label{sec:results}Results}
We set $a$ = 1 in this section, except in subsection~\ref{sec:ren}.
\subsection{Binding energies and meson masses}
In NRQCD, the heavy-light meson mass differs from the exponential falloff of 
the meson correlators, $E_{\rm sim}$, by the mass shift:
\beq
\Delta_{HL} = M - E_{\rm sim},
\eeq
 which depends on the renormalised 
heavy quark mass and the zero point of the non-relativistic  heavy quark 
energy, $E_0$:
\beq
\Delta_{HL} = Z_m \mnod - E_0.
\eeq
$Z_m$ is the heavy quark mass renormalisation constant. The 
meson mass can be determined nonperturbatively
 from the ratio of finite momentum and  zero momentum correlation 
functions~\cite{mackenzie}.
The dispersion relation of the heavy meson is given by:
\begin{equation}
E_{\rm sim}(\vec{p}) - E_{\rm sim}(0) = \sqrt{\vec{p}^2 + M^2} - M.
\end{equation}
We use the non-relativistic expansion of this:
\begin{equation}
E_{\rm sim}(\vec{p}) = E_{\rm sim}(0) + \frac{\vec{p}^2}{2M},
\end{equation}
where $M$ is the meson mass we want to determine. In this study the errors on the finite 
momentum correlators are rather large and we resort to different methods to calculate the
meson mass. The 
shift can also be determined perturbatively~\cite{colinII},
or nonperturbatively from the mass shift $\Delta_{HH}$ (defined analogous to $\Delta_{HL}$)
in  heavy-heavy systems:
\begin{equation}
\Delta_{HH} = 2\Delta_{HL}
\end{equation}
\begin{table}
\begin{center}
\begin{tabular}{|l|l|l|l|l|}
\hline
\multicolumn{1}{|c|}{} &
\multicolumn{2}{c|}{$\kappa_{\rm crit}$} &
\multicolumn{2}{c|}{$\kappa_{\rm s}(m_K)$} \\
\hline
\multicolumn{1}{|c|}{} &
\multicolumn{1}{|c|}{Run A} &
\multicolumn{1}{|c|}{Run B} &
\multicolumn{1}{|c|}{Run A} &
\multicolumn{1}{|c|}{Run B} \\
\hline
\multicolumn{1}{|c}{$\mnod$} &
\multicolumn{4}{|c|}{$E_{\rm sim}(PS)$} \\
\hline
1.71     & 0.437(19) &0.480(12)& 0.489(6) & 0.513(3) \\
2.0      & 0.437(11) &0.475(13)& 0.491(6) & 0.515(4) \\
2.5      & 0.439(12) &         & 0.496(7) &          \\
4.0      & 0.446(15) &0.489(16)& 0.499(6) & 0.528(4) \\
8.0      &           &0.482(16)&          & 0.527(6) \\
$\infty$ & 0.461(15) &         & 0.511(6) &  \\
\hline
\multicolumn{1}{|c}{$\mnod$} &
\multicolumn{4}{|c|}{$E_{\rm sim}(V)$} \\
\hline
1.71     & 0.464(15) & 0.498(15) & 0.509(6) & 0.532(4) \\
2.0      & 0.463(12) & 0.489(13) & 0.510(6) & 0.532(4) \\
2.5      & 0.464(9)  &           & 0.512(6) &          \\
4.0      & 0.478(12) & 0.505(15) & 0.515(6) & 0.539(4) \\
8.0      &           & 0.488(17) &          & 0.532(6) \\
\hline
\multicolumn{1}{|c}{$\mnod$} &
\multicolumn{4}{|c|}{$\overline{E_{\rm sim}}$} \\
\hline
1.71     & 0.457(14) & 0.493(10) & 0.504(6) &  0.528(3) \\
2.0      & 0.456(12) & 0.486(13) & 0.505(6) &  0.527(4) \\
2.5      & 0.457(9)  &           & 0.508(6) &           \\
4.0      & 0.462(13) & 0.501(13) & 0.511(6) &  0.536(3) \\
8.0      &           & 0.487(18) &          &  0.530(5) \\
\hline
\end{tabular}
\end{center}
\caption{Bare ground state energies from Run A and Run B at the 
strange and the chirally extrapolated light quark mass, in lattice
units.}
\label{tab:Esim_tadpole}
\end{table}
In Run B, the mass shifts for $\mnod$ = 1.71 and 2.0 are taken from the heavy-heavy results in
Ref.~\cite{upsilon},  the shift for $\mnod$ = 4.0 is  from a heavy-heavy 
simulation which was part of this project. For $\mnod$ = 8.0 we use perturbation theory since 
we expect the discretisation errors in heavy-heavy NRQCD for such large $\mnod$ to be  large.
The determination of the binding energies and the meson masses from Run A is discussed in 
Ref.~\cite{us_quenched}. 

\begin{table}
\begin{center}
\begin{tabular}{|l|c|c|}
\hline
\multicolumn{1}{|c|}{$\mnod$} &
\multicolumn{1}{c|}{$\Delta_{HL}$} &
\multicolumn{1}{c|}{$M_{PS}$} \\
\hline
1.71 & 1.73(10) & 2.21(11) \\
2.0 & 2.02(9) &  2.50(10) \\
4.0 & 4.07(5) &  4.56(7) \\
8.0 & 7.63(16) &  8.11(18) \\
\hline
\end{tabular}
\end{center}
\caption{Mass shifts used to calculate the meson masses in Run B, and chirally
extrapolated meson masses using these shifts. The error on the perturbative 
shift at $\mnod = 8.0$ is an estimate of the contributions from higher orders
in perturbation theory, obtained by squaring the one-loop contributions to
$E_0$ and $Z_m$. All numbers are in lattice units.}
\label{tab:shifts}
\end{table}

As shown in Table~\ref{tab:Esim_tadpole}, at $\kappa_{\rm crit}$ the pseudoscalar binding 
energies from Run A are $\sim\!10\%$ lower than from Run B, the vector $E_{\rm sim}$ are
$\sim\!5\%$ lower and the spin averaged $\overline{E_{\rm sim}}$, $\sim\!6-8\%$ lower. 
At $\kappa_{\rm s}$, all the
binding energies are $\sim\!4-5\%$ lower than the corresponding values from Run B. This 
difference corresponds to $\sim\!4\sigma$. In the chirally extrapolated case error bars are
larger and the difference between the two runs amounts to $2-3\sigma$. We expect $\Delta_{HL}$
to entirely depend on the heavy quark parameters~\cite{us_quenched}, thus the difference in 
$E_{\rm sim}$ translates directly into a difference in the meson mass. The values for 
$\Delta_{HL}$ used in Run B and the corresponding meson masses with chirally extrapolated
light quarks are given in Table~\ref{tab:shifts}. We note that the
pseudoscalar and spin averaged masses from Run B seem to be enhanced by at least as much as
the vector meson masses, which means that the enhancement of the clover term by
tadpole improvement does not increase the hyperfine splitting by an amount
greater than the statistical error. We came to the same conclusion
when we extracted the hyperfine splitting from the ratio of the $^3S_1$ and $^1S_0$ correlators (see Refs.~\cite{us_quenched,Melbourne}). Presumably the 
tadpole improvement of the light fermion affects the discretisation errors
in the kinetic energy and the spin-magnetic energy of the heavy quark by the
same fractional amount. Since the kinetic energy is much larger than the
spin-magnetic energy, the spin-independent part of $E_{\rm sim}$ receives the
larger absolute shift due to tadpole improvement.
\subsection{Bare lattice decay matrix elements}
In this section we discuss the bare, unrenormalised lattice matrix elements.  
Renormalisation constants will be dealt with in subsection~\ref{sec:ren}.

The $1/M$ corrections in NRQCD can be
separated into contributions of the kinetic and magnetic operator 
in the  Lagrangian and the correction to the local current:
\begin{equation}
f\sqrt{M} = (f\sqrt{M})^\infty\left( 1 + \frac{G_{kin}}{M} + 
\frac{2d_MG_{hyp}}{M} +
\frac{d_MG_{corr}/6}{M}\right). \label{eq:neubert}
\end{equation}
The notation is chosen to be consistent with Refs.~\cite{Neubert,Sara}.
For the axial current, $d_M$ = 3 and for the vector current, $d_M$ = $-1$.
With $(f\sqrt{M})^\infty$ we denote the static matrix element.

\subsubsection{\boldmath$f\protect\sqrt{M}$ from Run A}
Decay matrix elements in lattice units from Run A, 
for the pseudoscalar, vector and spin averaged cases, are shown in 
Table~\ref{table:fsqrtM}. To obtain an estimate of the physical matrix
elements and to be able to compare with other methods, we  chirally 
extrapolate in the light quark mass to $\kappa_{\rm crit}$ and interpolate to 
the strange light quark mass at $\kappa_{\rm s}$. These extrapolations should however
be used with some caution, because we only have two $\kappa$
values for the light fermions. For comparison, we also list results which do 
not include the corrections to the current, denoted as $(f\sqrt{M})^{uncorr}$.
The current corrections from Run A are given in Table~\ref{table:deltafsqrtM_A}.
\begin{table}
\begin{center}
\begin{tabular}{|l|l|c|l|l|}
\hline
                   \multicolumn{1}{|l|}{}
                  & \multicolumn{1}{l|}{$\kappa=0.1432$}
                 & \multicolumn{1}{l|}{$\kappa=0.1440$} 
                 & \multicolumn{1}{l|}{$\kappa_{\rm crit}$}
                 & \multicolumn{1}{l|}{$\kappa_{\rm s}(m_K)$} \\
\hline
\multicolumn{1}{|c}{$m_Q^{(0)}$} &
\multicolumn{4}{|c|}{$(f\sqrt{M})_{PS}^{uncorr}$} \\
\hline
 1.71    & 0.209(8) & 0.193(8) & 0.160(18) & 0.199(7) \\
 2.0     & 0.214(8) & 0.198(9) & 0.167(19) & 0.205(8) \\
2.5      & 0.226(7) & 0.206(10)& 0.17(2)   & 0.214(9) \\
4.0      & 0.249(11)& 0.227(13)& 0.19(3)   & 0.235(12)\\
$\infty$ & 0.332(7)& 0.315(9)  & 0.281(19) & 0.321(7)\\
\hline
\multicolumn{1}{|c}{$m_Q^{(0)}$} &
\multicolumn{4}{|c|}{$(f\sqrt{M})_{PS} $} \\
\hline
 1.71    & 0.165(7) & 0.153(8) & 0.128(21)  & 0.157(6) \\
 2.0     & 0.178(7) & 0.162(8) & 0.132(17)& 0.168(7)\\
2.5      & 0.195(7) & 0.176(9) & 0.138(19)& 0.183(8)\\
4.0      & 0.222(13)& 0.202(11)& 0.163(26)& 0.209(11)\\
\hline
\multicolumn{1}{|c}{$m_Q^{(0)}$} &
\multicolumn{4}{|c|}{$(f\sqrt{M})_{V}^{uncorr}$} \\
\hline
 1.71   & 0.195(10)& 0.180(9) & 0.15(2)   & 0.186(8)  \\
 2.0    & 0.202(7) & 0.186(7) & 0.155(16) & 0.192(7)  \\
2.5     & 0.215(9) & 0.197(8) & 0.161(13) & 0.204(8)  \\
4.0     & 0.244(8) & 0.224(11)& 0.185(17) & 0.233(10) \\
\hline
\multicolumn{1}{|c}{$m_Q^{(0)}$} &
\multicolumn{4}{|c|}{$(f\sqrt{M})_{V} $} \\
\hline
 1.71  & 0.223(9)& 0.205(9)    & 0.171(20) & 0.212(8) \\
 2.0    & 0.226(8) & 0.210(9)  & 0.177(19) & 0.216(8) \\
2.5     & 0.237(8) & 0.216(11) & 0.176(24) & 0.224(10)\\
4.0     & 0.256(11)& 0.234(13) & 0.191(24) & 0.242(12)\\
\hline
\multicolumn{1}{|c}{$m_Q^{(0)}$} &
\multicolumn{4}{|c|}{$\overline{f\sqrt{M}} $} \\
\hline
 1.71   & 0.199(9) & 0.183(8) & 0.153(18) & 0.189(7) \\
 2.0    & 0.205(6) & 0.189(7) & 0.158(14) & 0.195(6) \\
2.5     & 0.218(8) & 0.199(7) & 0.163(11) & 0.206(8) \\
4.0     & 0.245(8) & 0.225(11)& 0.185(18) & 0.233(10) \\
\hline
\end{tabular}
\end{center}
\caption{Decay matrix elements $f\protect\sqrt{M}$ in lattice units
from Run A. }
\label{table:fsqrtM}
\end{table}
\begin{table}[h]
\begin{center}
\begin{tabular}{|c|l|c|l|l|}
\hline
                   \multicolumn{1}{|l|}{}
                  & \multicolumn{1}{l|}{$\kappa=0.1432$}
                 & \multicolumn{1}{l|}{$\kappa=0.1440$} 
                 & \multicolumn{1}{l|}{$\kappa_{\rm crit}$}
                 & \multicolumn{1}{l|}{$\kappa_{\rm s}(m_K)$} \\
\hline	    	      
\multicolumn{1}{|c}{$\mnod$} &
\multicolumn{4}{|c|}{$\delta(f\sqrt{M})_{PS}$} \\
\hline                              
1.71   & $-0.0412(16)$  & $-0.0383(17)$ & $-0.033(4)$  & $-0.0394(14)$ \\
2.0    & $-0.0364(13)$  & $-0.0340(15)$ & $-0.029(3)$  & $-0.0349(14)$ \\
2.5    & $-0.0312(10)$  & $-0.0287(14)$ & $-0.024(4)$  & $-0.0297(13)$ \\
4.0    & $-0.0219(9) $  & $-0.0202(11)$ & $-0.017(2)$  & $-0.0208(11)$ \\
\hline	    	      
\end{tabular}
\end{center}
\caption{Current corrections to the decay matrix elements from Run A, in lattice units.}
\label{table:deltafsqrtM_A}
\end{table}
\begin{figure}[pthb]         
\vspace{-2cm}
\centerline{\epsfxsize=13cm
\epsfbox{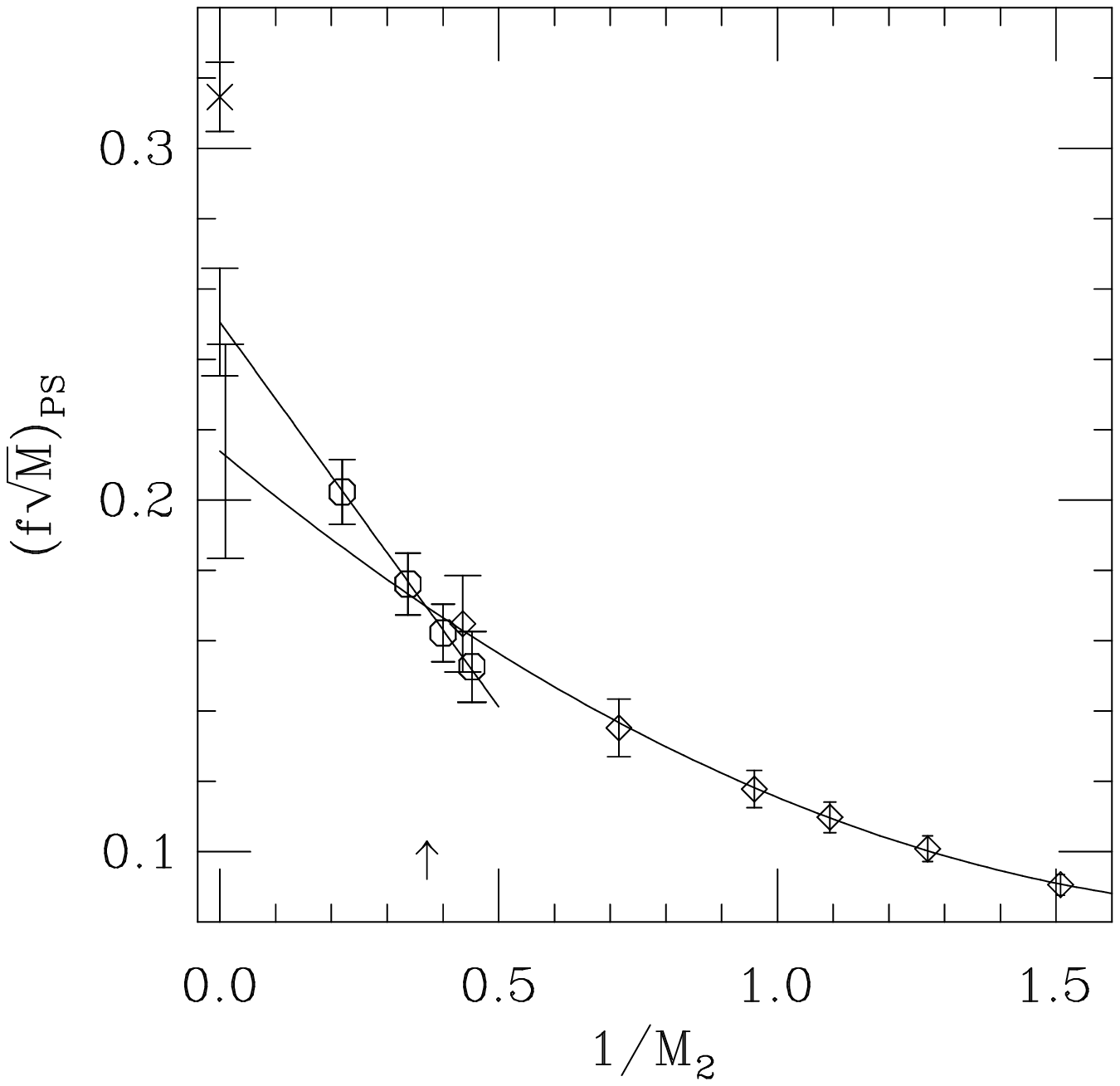}
}
\caption{Unrenormalised matrix elements $(f\protect\sqrt{M})_{PS}$ in lattice 
units with NRQCD heavy quarks from Run 
A (circles) and with tree-level clover heavy quarks 
(diamonds)~\protect\cite{kappac,sara_phd} at $\kappa = 
0.1440$. The cross denotes the static point from Run A. 
The lines are correlated  fits; the errors on the 
extrapolations to infinite mass are for clarity slightly shifted from the
origin. The arrow denotes the position of the $B_s$ meson in this plot (the
light quark mass at $\kappa = 0.1440$ is close to the strange quark mass).}
\label{fig:UKQCD}
\end{figure}
On the same set of configurations, the matrix elements have been calculated 
using clover heavy quarks ($c_{SW} = 1$) by the UKQCD 
Collaboration~\cite{kappac,sara_phd}. Thus we are able to make a direct
comparison between NRQCD and clover heavy quarks. Fig.~\ref{fig:UKQCD} shows
$(f\sqrt{M})_{PS}$ as a function of the inverse pseudoscalar meson mass 
$M_{PS}$, both for
clover and NRQCD (with the current correction included) heavy quarks. 
\begin{table}[thpb]
\begin{center}
\begin{tabular}{|c|l|c|l|l|}
\hline
                   \multicolumn{1}{|c|}{$t_{min}/t_{max}$}
                 & \multicolumn{1}{l|}{$\kappa=0.1432$}
                 & \multicolumn{1}{l|}{$\kappa=0.1440$} 
                 & \multicolumn{1}{l|}{$\kappa_{\rm crit}$}
                 & \multicolumn{1}{l|}{$\kappa_{\rm s}$} \\
\hline
3/10 & 0.333(6) & 0.319(6) & 0.293(8) & 0.324(5)\\
4/10 & 0.334(7) & 0.317(7) & 0.285(9) & 0.323(7)\\
5/10 & 0.333(9) & 0.316(9) & 0.282(15)& 0.322(8)\\
6/10 & 0.337(10)& 0.319(13)& 0.29(4)  & 0.326(11) \\
7/10 & 0.343(13)& 0.324(13)& 0.29(3)  & 0.331(14) \\
\hline
\end{tabular}
\end{center}
\caption{The static  $f\protect\sqrt{M}$ from Run A
for various fit intervals, in lattice units.}
\label{table:static_vs_t}
\end{table}
\begin{figure}[tphb]
\vspace{-2cm}
\centerline{
\epsfxsize=8cm
\hspace{1cm}\epsfbox{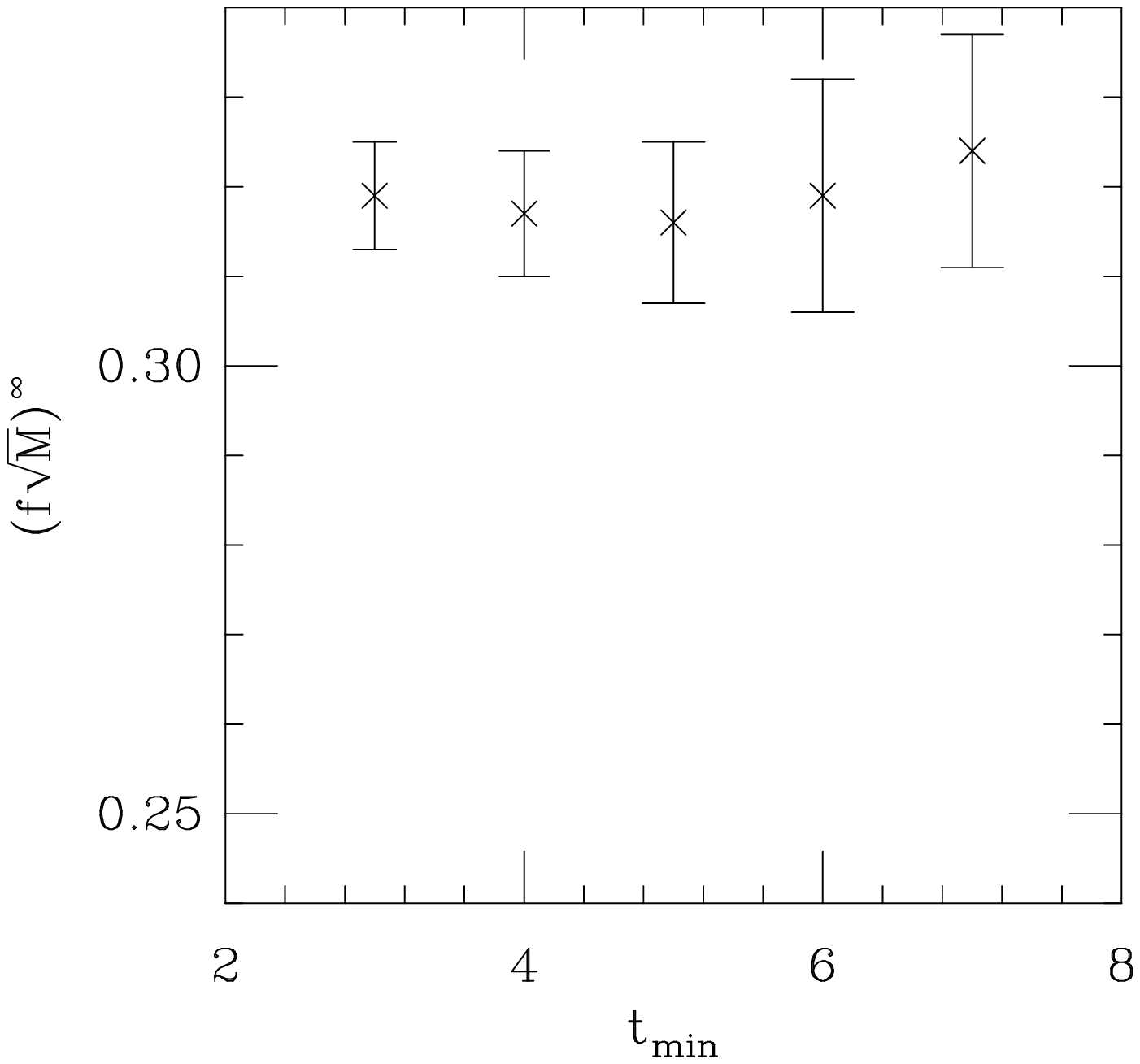}
\epsfysize=8cm
\hspace{-0.8cm}\epsfbox{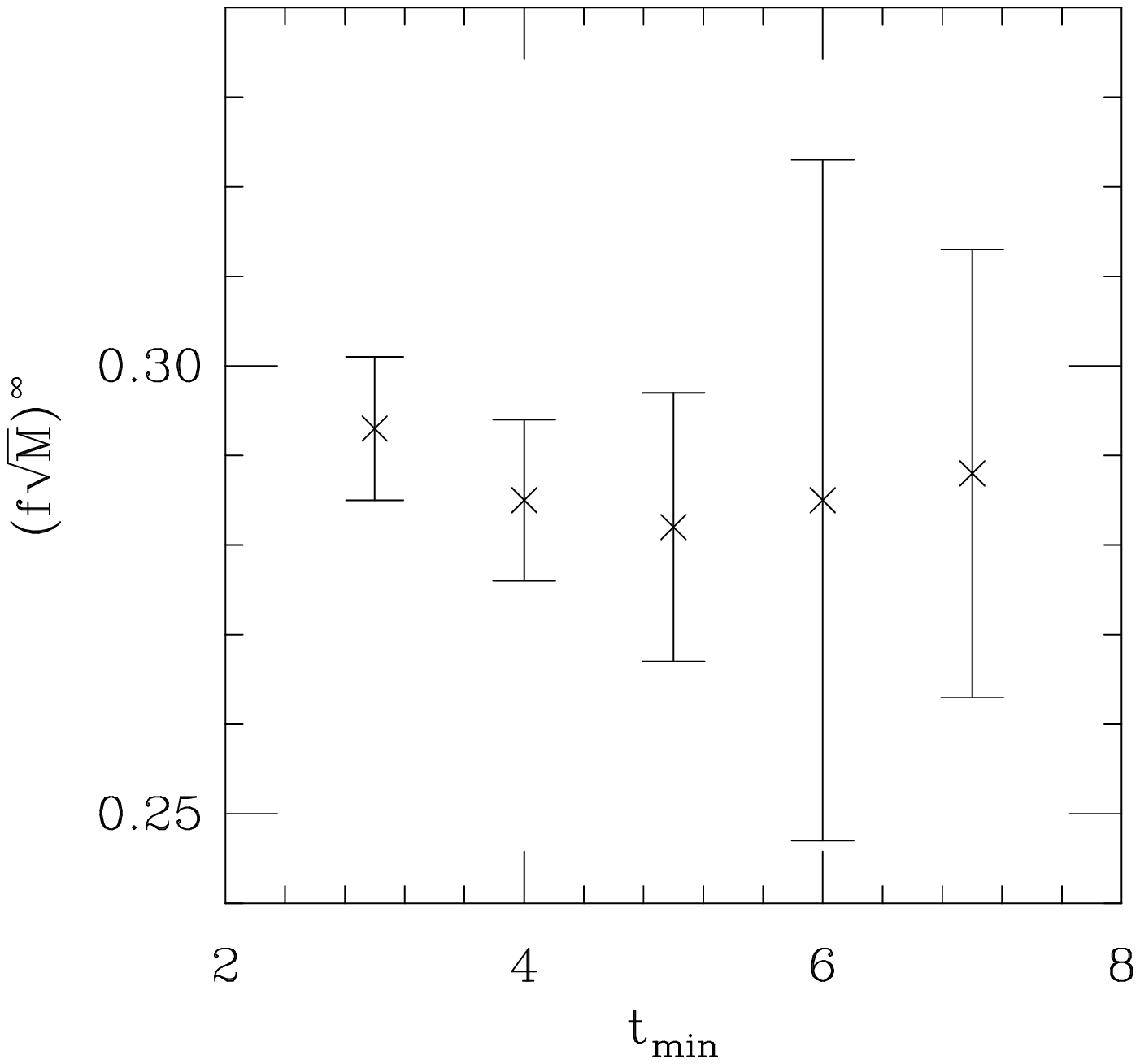}
}
\caption{Dependence of the static matrix element from Run A, in lattice units,
 on the fit interval.
On the left, $\kappa = 0.1440$, on the right, $\kappa = \kappa_{\rm crit}$.}
\label{fig:sta_vs_tmin}
\end{figure}
\begin{figure}[bhtp]
\vspace{-2cm}
\centerline{
\epsfxsize=8cm
\hspace{1cm}\epsfbox{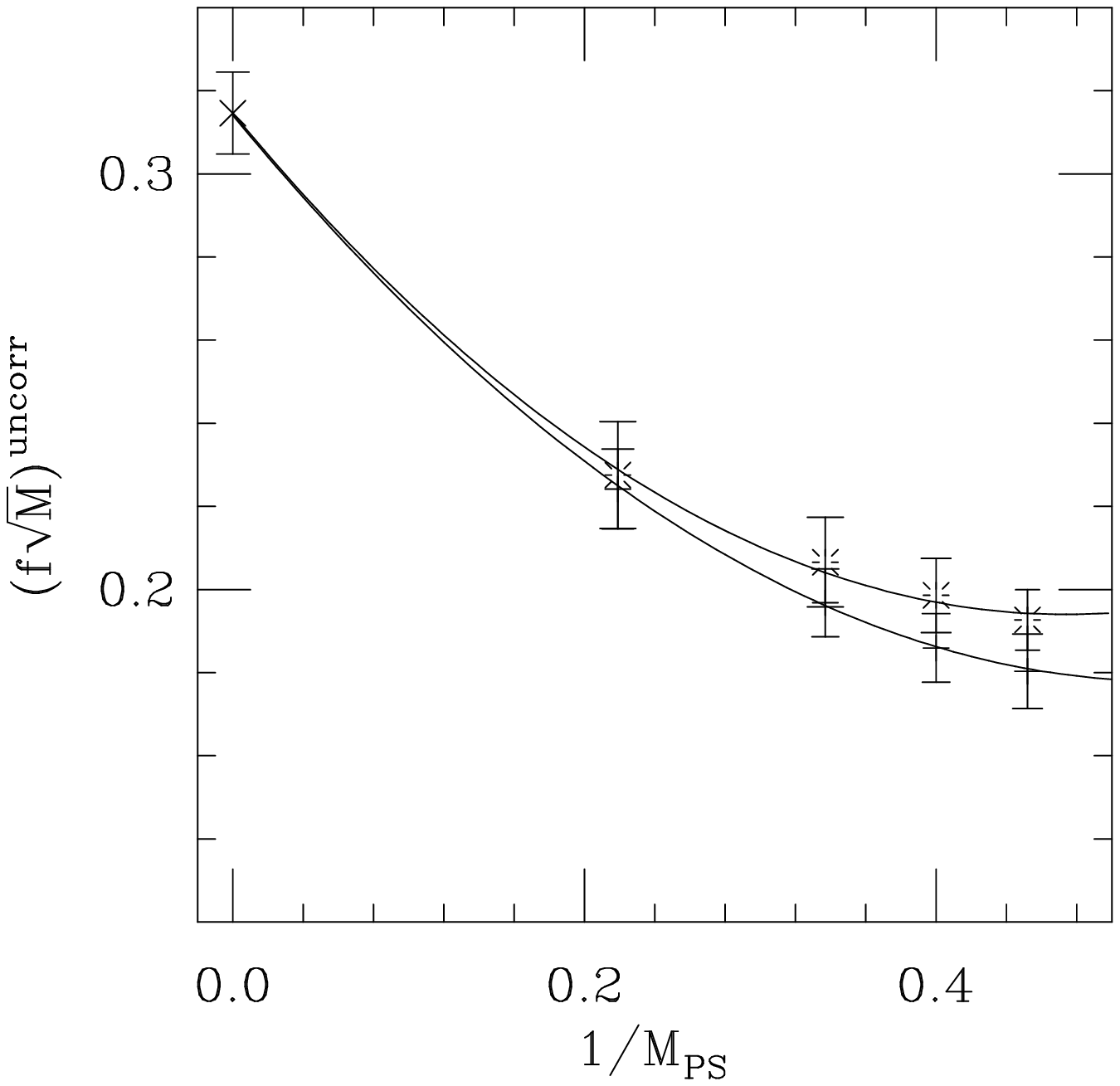}
\epsfxsize=8cm
\hspace{-0.8cm}\epsfbox{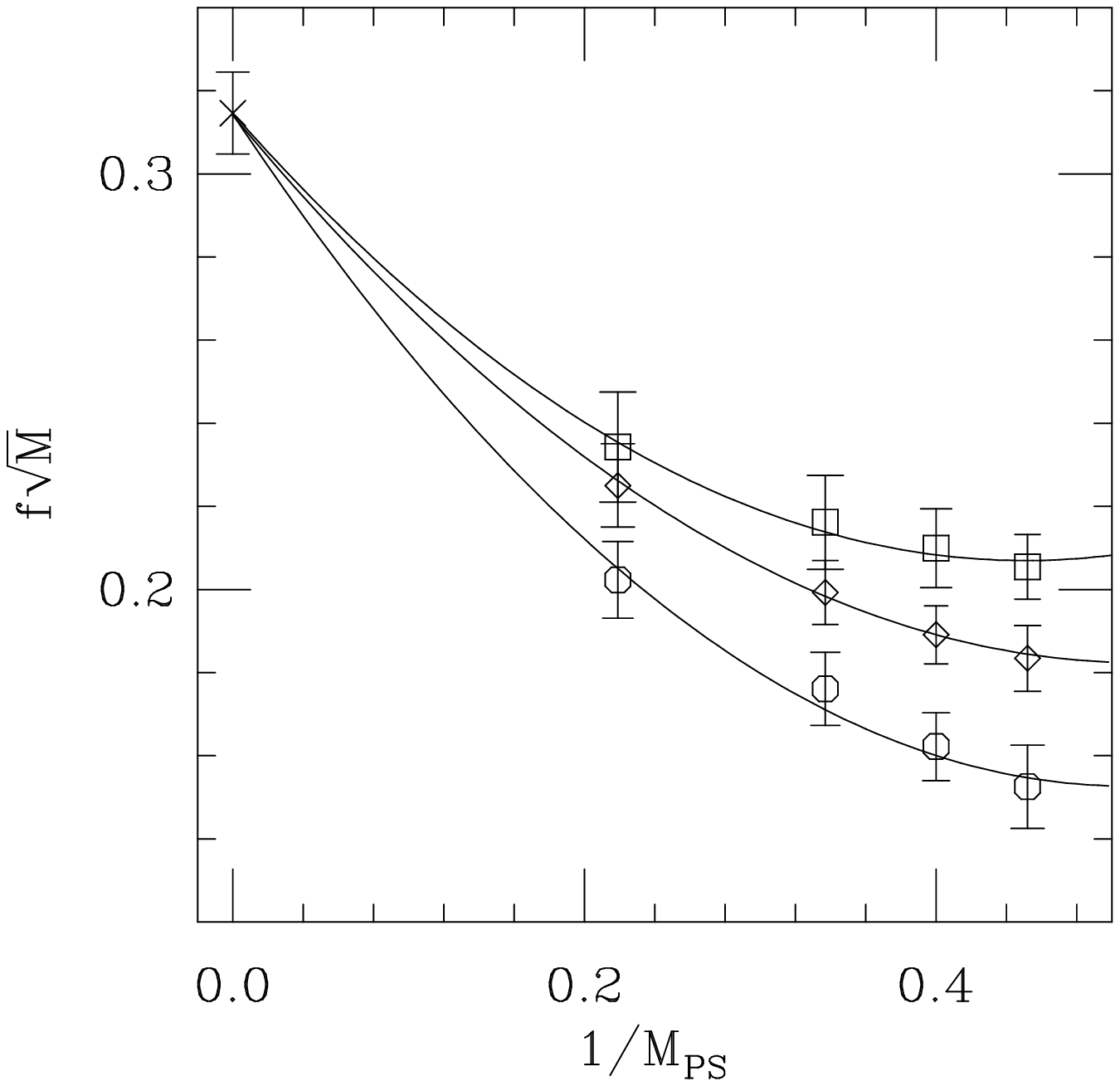}
}
\caption{Unrenormalised decay matrix elements from Run A, in lattice units. 
On the left, pseudoscalar (bursts)
and vector (pluses) matrix elements without the current correction included.
On the right, pseudoscalar (circles), and vector (squares) decay constants 
with the current corrections. The diamonds on the right denote the 
spin-averaged case. The lines are correlated fits including the static point  
(cross). $\kappa$ = 0.1440.}
\label{fig:fsqrtM_A}
\end{figure}
Note that for the mesons with clover heavy quarks we use
$M_2$,  the `dynamical'  meson mass 
determined from the dispersion relation of the heavy meson~\cite{sara_phd}.
The NRQCD results from Run A can be fit to a linear function in the inverse mass;
a possible quadratic dependence of these results on the heavy mass cannot be 
resolved. 
The bare lattice matrix elements from both types of heavy quarks agree in 
the $B$ region within errors. This is not necessarily true for the renormalised
matrix elements. Renormalisation decreases $(f\sqrt{M})_{PS}$ from NRQCD in 
the $B$
region by $\sim\!10\%$, as detailed in subsection~\ref{sec:ren}, whereas
the current renormalisation constants have not been calculated for clover 
quarks with large mass.

Neither of the results presented in 
Fig.~\ref{fig:UKQCD} extrapolates in the infinite mass limit to the static
simulation result. 
Our static point agrees with UKQCD  using different (Jacobi) 
smearing~\cite{kappac}, and it appears that there is no sizeable  excited state
contamination which could artificially enhance the static result 
(see Table~\ref{table:static_vs_t} and Fig.~\ref{fig:sta_vs_tmin}). Instead, 
the reason seems to be that 
the quality of the results from Run A does not allow us to predict the slope 
at infinite quark mass correctly. In NRQCD calculations which include larger 
masses and better ensembles the results do extrapolate to the static 
point. This is detailed in the discussion of Run B in this paper (see 
Fig.~\ref{fig:fALL_3700}) and in
Refs.~\cite{Sara,saraprep}). In these studies we find that in the region of 
the $B$ there is a quadratic contribution to the slope of $f\sqrt{M}$.
We do however obtain good quadratic fits of the results from Run A if we 
include the static point into the fit.
In Figure~\ref{fig:fsqrtM_A} we show the NRQCD and static decay matrix elements from 
Run A at $\kappa$ = 0.1440 with a correlated fit  to a 
quadratic function in the inverse meson mass that includes the static point. 
For the  results without 
inclusion of the current corrections, at the left in Fig.~\ref{fig:fsqrtM_A}, 
we find $Q = 0.94$ for the pseudoscalar and $Q = 0.96$ for the vector matrix element.
The vector matrix element is smaller than the pseudoscalar
matrix element. On the right, we plot the matrix elements with the correction to the
current. The current correction to $(f\sqrt{M})_V$ has not been simulated
in Run A, but we  estimated the corrected vector current (see Table~\ref{table:fsqrtM})
 using the axial current corrections in Table~\ref{table:fsqrtM} and the
relation:
\beq
\delta(f\sqrt{M})_V = -\frac{1}{3}\frac{\delta(f\sqrt{M})_{PS}}{(f\sqrt{M})_{PS}^{uncorr}}
\cdot(f\sqrt{M})_{PS}^{uncorr},
\eeq
which follows from Eq.~(\ref{eq:neubert}) with the appropriate values for $d_M$.
We find  $(f\sqrt{M})_{PS}$ in the $B$ region to be much lower, around 50\%,
than in the static case. The current correction 
gives a contribution of relative size 22 \% in the $B$ region.  
The spin averaged matrix element, defined as:
\begin{equation}
\overline{f\sqrt{M}} = \frac{1}{4}\left[(f\sqrt{M})_{PS} + 3(f\sqrt{M})_V\right],
\end{equation}
is also shown in Fig.~\ref{fig:fsqrtM_A}.
It can be calculated from the uncorrected matrix elements, since the current correction
drops out after spin averaging (see Eq.~(\ref{eq:neubert})). With the current corrections
included, the quadratic fit of the pseudoscalar matrix element gives $Q=0.58$, of the 
vector matrix element $Q = 0.95$, and the spin averaged, $Q = 0.93$.
\subsubsection{\boldmath$f\protect\sqrt{M}$ from Run B}
\begin{figure}[pthb]         
\vspace{-2cm}
\centerline{\epsfxsize=13cm
\epsfbox{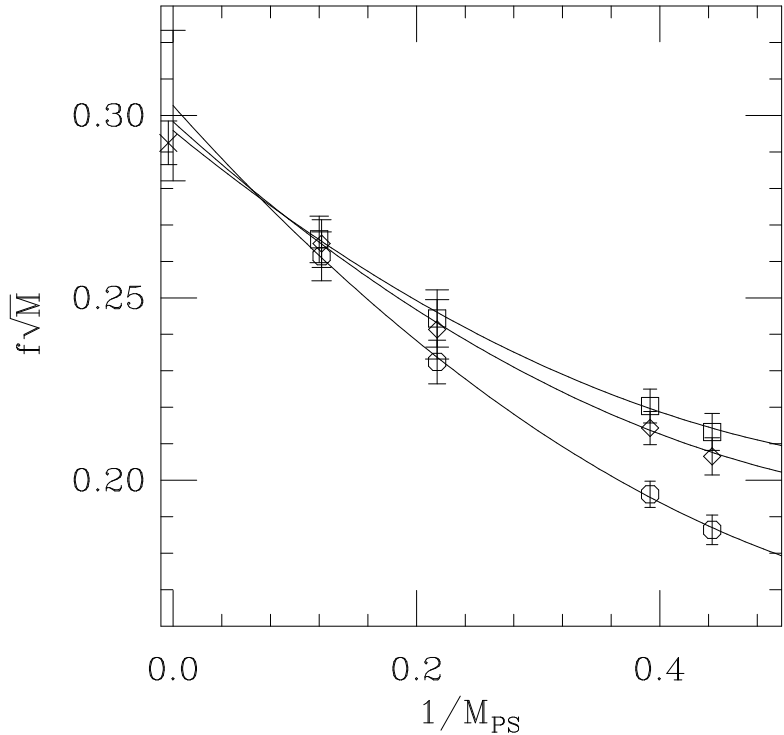}
}
\caption{Unrenormalised decay matrix elements from Run B at $\kappa$ = 0.1370 plotted against the
inverse pseudoscalar meson mass. Circles denote $(f\protect\sqrt{M})_{PS}$,
squares $(f\protect\sqrt{M})_{V}$, and diamonds the spin-averaged matrix
element. Lines denote correlated fits; we indicate the error bar on the extrapolation
of the pseudoscalar matrix element to infinite mass. The static point
(cross) is slightly shifted from the origin for clarity. All quantities are in 
lattice units.}
\label{fig:fALL_3700}
\end{figure}
\begin{figure}[pthb]         
\vspace{-2cm}
\centerline{\epsfxsize=13cm
\epsfbox{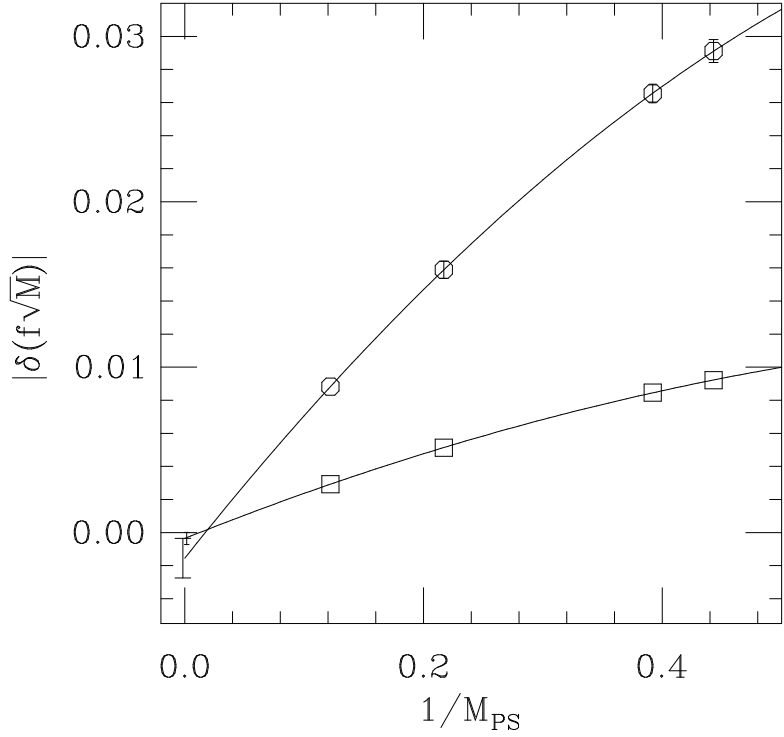}
}
\caption{Unrenormalised current corrections from Run B, in lattice units,
 at $\kappa$ = 0.1370, plotted against the
inverse pseudoscalar meson mass. Circles denote $-\delta(f\protect\sqrt{M})_{PS}$,
and squares $\delta(f\protect\sqrt{M})_{V}$. The lines denote correlated fits of
$|\delta(f\protect\sqrt{M})|$ to a quadratic function in $1/M_{PS}$. Where not
shown, error bars are smaller than the symbols.}
\label{fig:deltafP_3700}
\end{figure}
\begin{table}
\begin{center}
\begin{tabular}{|l|l|l|l|l|}
\hline
\multicolumn{1}{|c|}{} &
\multicolumn{1}{c|}{$\kappa$ = 0.1370} &
\multicolumn{1}{c|}{$\kappa$ = 0.1381} &
\multicolumn{1}{c|}{$\kappa_{\rm crit}$} &
\multicolumn{1}{c|}{$\kappa_{\rm s}(m_K)$} \\
\hline
\multicolumn{1}{|c}{$m_Q^{(0)}$} &
\multicolumn{4}{|c|}{$(f\sqrt{M})_{PS}^{uncorr}$} \\
\hline
1.71     & 0.212(5) & 0.201(6)& 0.188(14) & 0.206(4) \\
2.0      & 0.219(5) & 0.203(4)& 0.187(12) & 0.211(3)\\
4.0      & 0.242(8) & 0.225(6)& 0.206(15) & 0.233(4) \\
8.0      & 0.262(9) & 0.243(8)& 0.223(18) & 0.252(6) \\
$\infty$ & 0.293(6) & 0.274(7) & 0.254(15)&0.283(5) \\
\hline
\multicolumn{1}{|c}{$m_Q^{(0)}$} &
\multicolumn{4}{|c|}{$(f\sqrt{M})_{PS} $} \\
\hline
1.71     & 0.186(4) &0.173(5) & 0.158(12) & 0.179(3)\\
2.0      & 0.196(4) &0.178(4) & 0.158(9)  & 0.186(3)\\
4.0      & 0.232(6) &0.210(6) & 0.186(14) & 0.221(4)\\
8.0      & 0.261(7) &0.237(9) & 0.211(17) & 0.248(5)\\
\hline
\multicolumn{1}{|c}{$m_Q^{(0)}$} &
\multicolumn{4}{|c|}{$(f\sqrt{M})_{V}^{uncorr}$} \\
\hline
1.71     &0.202(5) & 0.194(9) & 0.185(18) & 0.198(5) \\
2.0      &0.210(5) & 0.192(6) & 0.174(12) & 0.201(4) \\
4.0      &0.236(9) & 0.217(6) & 0.197(16) & 0.226(5) \\
8.0      &0.261(6) & 0.240(9) & 0.218(19) & 0.250(6) \\
\hline
\multicolumn{1}{|c}{$m_Q^{(0)}$} &
\multicolumn{4}{|c|}{$(f\sqrt{M})_{V} $} \\
\hline
1.71     & 0.213(5) & 0.198(6) & 0.181(16) & 0.205(4)  \\
2.0      & 0.220(5) & 0.200(7) & 0.179(14) & 0.210(4)  \\
4.0      & 0.244(8) & 0.226(7) & 0.206(16) & 0.235(5)  \\
8.0      & 0.266(8) & 0.243(8) & 0.219(17) & 0.254(6)  \\
\hline
\multicolumn{1}{|c}{$m_Q^{(0)}$} &
\multicolumn{4}{|c|}{$\overline{f\sqrt{M}} $} \\
\hline
1.71     & 0.207(5) & 0.191(5) & 0.175(12) & 0.199(4) \\
2.0      & 0.214(5) & 0.195(6) & 0.174(12) & 0.204(4) \\
4.0      & 0.241(7) & 0.222(7) & 0.201(13) & 0.231(4) \\
8.0      & 0.265(7) & 0.242(10)& 0.217(18) & 0.253(6) \\
\hline
\end{tabular}
\end{center}
\caption{Decay matrix elements  from Run B, in lattice units.}
\label{tab:fsqrtMtadpole1}
\end{table}
\begin{table}
\begin{center}
\begin{tabular}{|l|l|l|l|l|}
\hline                        
\multicolumn{1}{|c|}{} &
\multicolumn{1}{c|}{$\kappa$ = 0.1370} &
\multicolumn{1}{c|}{$\kappa$ = 0.1381} &
\multicolumn{1}{c|}{$\kappa_{\rm crit}$} &
\multicolumn{1}{c|}{$\kappa_{\rm s}(m_K)$} \\
\hline
\multicolumn{1}{|c}{$\mnod$} &
\multicolumn{4}{|c|}{$\delta(f\sqrt{M})_{PS}$} \\
\hline
1.71 & $-0.0291(7)$ & $-0.0287(9)$& $-0.028(2)$   & $-0.0289(5)$ \\
2.0  & $-0.0265(6)$ & $-0.0252(6)$& $-0.0238(16)$ & $-0.0258(4)$ \\
4.0  & $-0.0159(5)$ & $-0.0151(4)$& $-0.0143(10)$ & $-0.0155(3)$ \\
8.0  & $-0.0088(3)$ & $-0.0084(3)$& $-0.0079(6)$  & $-0.0086(2)$ \\
\hline                        
\multicolumn{1}{|c}{$\mnod$} &
\multicolumn{4}{|c|}{$\delta(f\sqrt{M})_{V}$} \\
\hline
1.71 & $\hphantom{-}0.0092(2)$   & $\hphantom{-}0.0091(4)$   & 
$\hphantom{-}0.089(9)$ & $\hphantom{-}0.091(3)$   \\
2.0  & $\hphantom{-}0.0085(2)$   & $\hphantom{-}0.0079(2)$   & 
$\hphantom{-}0.073(5)$ & $\hphantom{-}0.0816(14)$ \\
4.0  & $\hphantom{-}0.00513(18)$ & $\hphantom{-}0.00485(13)$ & 
$\hphantom{-}0.045(4)$ & $\hphantom{-}0.0498(11)$ \\
8.0  & $\hphantom{-}0.00292(7)$  & $\hphantom{-}0.00277(10)$ & 
$\hphantom{-}0.026(2)$ & $\hphantom{-}0.0284(7)$  \\
\hline
\end{tabular}
\end{center}
\caption{Current corrections to the decay matrix elements from Run B, in lattice units.}
\label{table:deltafsqrtM_B}
\end{table}

In Fig.~\ref{fig:fALL_3700} we show the pseudoscalar, vector and spin averaged decay matrix 
element from Run B with correlated fits of the bare simulation results to  a quadratic
function in the inverse  pseudoscalar meson mass. The results of these fits
for the pseudoscalar matrix elements for both light $\kappa$ values  are 
shown in Table~\ref{tab:fit_fsqrtM}. 
\begin{table}
\begin{center}
\begin{tabular}{|c|c|c|c|}
\hline
\multicolumn{1}{|c|}{$\kappa$} &
\multicolumn{1}{c|}{$a_1$} &
\multicolumn{1}{c|}{$a_2/a_1$} &
\multicolumn{1}{c|}{$a_3/a_1$} \\
\hline
$0.1370$ & 0.30(2) & $-1.2(4)$ & 1.6(1.5) \\
$0.1381$ & 0.28(3) & $-1.4(7)$ & 2.6(2.4) \\
\hline
\end{tabular}
\end{center}
\caption{Results of correlated fits of $(f\protect\sqrt{M})_{PS}$ in
lattice units
from Run B to the function $a_1+a_2/M_{PS}+a_3/M_{PS}^2$.}
\label{tab:fit_fsqrtM}
\end{table}
\begin{table}
\begin{center}
\begin{tabular}{|c|c|c|c|}
\hline
\multicolumn{1}{|c|}{} &
\multicolumn{1}{l|}{$a_1$} &
\multicolumn{1}{l|}{$a_2$} &
\multicolumn{1}{l|}{$a_3$} \\
\hline
\multicolumn{1}{|c|}{$\kappa$} &
\multicolumn{3}{|c|}{$\delta(f\sqrt{M})_{PS}$} \\
\hline
$0.1370$ &$-0.0016(13)$ & $0.091(11)$ & $-0.10(4)$ \\
$0.1381$ &$-0.0016(14)$ & $0.087(14)$ & $-0.09(4)$ \\
\hline
\multicolumn{1}{|c|}{$\kappa$} &
\multicolumn{3}{|c|}{$\delta(f\sqrt{M})_{V}$} \\
\hline
$0.1370$ &$-0.0016(13)$ & $0.091(11)$ & $-0.10(4)$ \\
$0.1381$ &$-0.0004(5)$ & $0.028(5)$ & $-0.033(15)$ \\
\hline
\end{tabular}
\end{center}
\caption{Result of correlated fits of the current corrections from Run B
 to the function $a_1 + 
a_2/M_{PS} + a_3/M_{PS}^2$. All quantities are in lattice units.}
\label{tab:deltas}
\end{table}
Within errors, the infinite mass limit of the NRQCD matrix elements are in agreement with the
static simulation result. We find the relative slope of $f\sqrt{M}$ to be of the order of 
2.5 GeV. The error on the slope is however large (see Table~\ref{tab:fit_fsqrtM}),
because of our small ensemble size, and
the number of degrees of freedom for these fits is low.
Our results from a simulation using dynamical configurations with 
higher  statistics~\cite{saraprep} indicate that the slope actually gets smaller with 
decreasing light quark mass. Renormalisation decreases the NRQCD decay matrix elements 
relative to the bare ones (see subsection~\ref{sec:ren}), but their infinite mass
limit is still in agreement with the (renormalised) static matrix elements.

To study the behaviour  of the lattice matrix elements for the current corrections in the 
infinite mass limit, we perform correlated fits of $\delta(f\protect\sqrt{M})_{PS}$ and
$\delta(f\protect\sqrt{M})_{V}$ as functions in the inverse pseudoscalar meson mass. Fits to
a linear function have $Q \ll 1$, but  fits
to a quadratic function work well. The
results are presented in Table~\ref{tab:deltas}.
We find that the infinite mass limit of our results is in a 
reasonable agreement with zero. 
In Fig.~\ref{fig:deltafP_3700}, the axial and vector current corrections for
$\kappa$ = 0.1370 are shown.
\begin{figure}[hbtp]         
\vspace{-2cm}
\centerline{\epsfxsize=13cm
\epsfbox{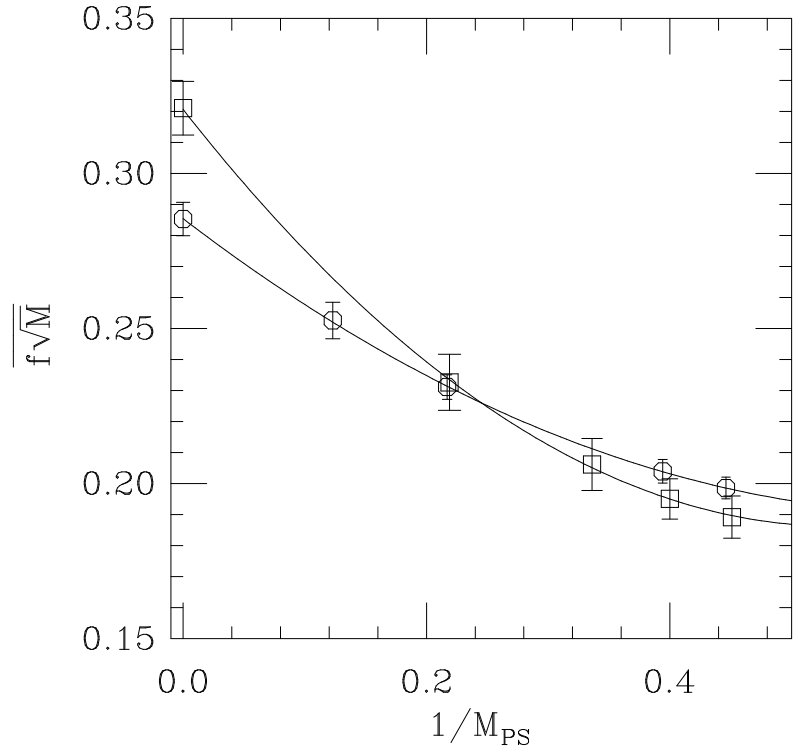}
}
\caption{Spin averaged decay matrix elements, unrenormalised, in lattice units,
 at $\kappa_{\rm s}$, plotted against the inverse pseudoscalar meson mass. Squares
are results from Run A, and circles, from Run B. Lines
denote correlated fits to all the points including the static.}
\label{fig:fAVG_kappas_both}
\end{figure}

From the heavy mass dependence of the spin average of the matrix elements, we 
can, using Eq.~(\ref{eq:neubert}), extract the contribution of the kinetic energy
 to $f\sqrt{M}$~\cite{Sara}:
\beq
\overline{f\sqrt{M}} = (f\sqrt{M})^\infty \left(1 + \frac{G_{kin}}{M}\right).
\eeq
We fit the 
spin-averaged matrix elements from Run A and Run B, including also the static
point, to a quadratic function in $1/M_{PS}$. Results in lattice units
for the strange and the chirally extrapolated light quark mass can be found in 
Table~\ref{tab:fit_Gkin}. Fig.~\ref{fig:fAVG_kappas_both} shows 
$\overline{f\sqrt{M}}$ at $\kappa_{\rm s}$ from both runs.
For Run B, $G_{kin}$ is found to be $\sim 2\sigma$
smaller than for Run A. However it has to be noted here that the fits include 
5 points and thus there are only 2 degrees of freedom. 
For a definite conclusion it would be necessary to include
more simulation results at heavy quark masses. Moreover, things are
expected to change after inclusion of renormalisation constants.
We expect the heavy mass dependence
for each run to change after renormalisation in a different way. In particular, it
appears that rotating 
the light quark in Run A with the $\gamma\cdot D$ operator  introduces  
an additional, heavy mass dependent, contribution to $Z_A$ and $Z_V$ in 
perturbation theory.
\begin{table}
\begin{center}
\begin{tabular}{|c|c|c|}
\hline
\multicolumn{1}{|c|}{} &
\multicolumn{1}{c|}{Run A} &
\multicolumn{1}{c|}{Run B} \\
\hline
$\kappa_{\rm s}$    & $-1.6(2)$  & $-1.05(15)$ \\
$\kappa_{\rm crit}$ & $-2.0(5)$  & $-1.2(5)$ \\
\hline
\end{tabular}
\end{center}
\caption{Results for $G_{kin}$ from correlated fits of 
$(\overline{f\protect\sqrt{M}})$ to a quadratic function in 
$1/M_{PS}$, in lattice units.}
\label{tab:fit_Gkin}
\end{table}
\subsubsection{\boldmath$(f\protect\sqrt{M})_{PS}/(f\protect\sqrt{M})_V$}
We can also study the behaviour of  the ratios of axial and vector matrix 
elements in the heavy quark limit. The ratio of the matrix elements without
the current correction should give an estimate for the 
contribution of the spin-magnetic interaction in the Hamiltonian to $f\sqrt{M}$ in 
Eq.~(\ref{eq:neubert})~\cite{Sara}:
\beq
R^{uncorr} \equiv \frac{(f\sqrt{M})_{PS}^{uncorr}}{(f\sqrt{M})_{V}^{uncorr}} 
\propto 1 + \frac{8G_{hyp}}{M}. 
\eeq
For the slope of the ratio of the matrix elements with the current correction one 
expects:
\beq
R \equiv \frac{(f\sqrt{M})_{PS}}{(f\sqrt{M})_{V}} \propto
1 + \frac{8G_{hyp}+2G_{corr}/3}{M}. \label{eq:hypcorr}
\eeq
\begin{table}
\begin{center}
\begin{tabular}{|l|c|c|c|c|}
\hline
\multicolumn{5}{|c|}{Run A} \\
\hline
                   \multicolumn{1}{|c|}{}
                 & \multicolumn{1}{c|}{$\kappa=0.1432$}
                 & \multicolumn{1}{c|}{$\kappa=0.1440$} 
                 & \multicolumn{1}{c|}{$\kappa_{\rm crit}$}
                 & \multicolumn{1}{c|}{$\kappa_{\rm s}$} \\
\hline
 \multicolumn{1}{|c|}{$m_Q^{(0)}$}
 & \multicolumn{4}{c|}{$(f\sqrt{M})^{uncorr}_{PS}/(f\sqrt{M})^{uncorr}_V$} \\
\hline
1.71 & 1.07(5) & 1.07(5) & 1.07(18) &  1.07(4) \\
2.0  & 1.06(4) & 1.07(4) & 1.09(14) &  1.07(3) \\
2.5  & 1.05(3) & 1.05(3) & 1.05(20) &  1.05(3) \\
4.0  & 1.02(3) & 1.02(5) & 1.02(18) &  1.02(4) \\
\hline
 \multicolumn{1}{|c|}{$\mnod$}
 & \multicolumn{4}{c|}{$(f\sqrt{M})_{PS}/(f\sqrt{M})_V$} \\
\hline
1.71 & 0.74(3)  & 0.74(3) & 0.75(13) & 0.74(2) \\
2.0  & 0.78(2)  & 0.77(3) & 0.76(14) & 0.78(2)  \\
2.5  & 0.826(16)& 0.82(3) & 0.80(12) & 0.82(2)  \\
4.0  & 0.87(4)  & 0.86(4) & 0.86(19) & 0.86(4)  \\
\hline
\multicolumn{5}{|c|}{Run B} \\
\hline
                   \multicolumn{1}{|c|}{}
                 & \multicolumn{1}{c|}{$\kappa=0.1370$}
                 & \multicolumn{1}{c|}{$\kappa=0.1381$} 
                 & \multicolumn{1}{c|}{$\kappa_{\rm crit}$}
                 & \multicolumn{1}{c|}{$\kappa_{\rm s}$} \\
\hline
 \multicolumn{1}{|c|}{$m_Q^{(0)}$}
 & \multicolumn{4}{c|}{$(f\sqrt{M})^{uncorr}_{PS}/(f\sqrt{M})^{uncorr}_V$} \\
\hline
1.71 & 1.05(2)   & 1.04(5)   & 1.03(12) & 1.04(3)  \\
2.0  & 1.041(16) & 1.057(19) & 1.08(5)  & 1.049(12) \\
4.0  & 1.028(19) & 1.036(18) & 1.05(5)  & 1.032(13)  \\
8.0  & 1.002(18) & 1.010(17) & 1.02(5)  & 1.006(13)  \\ 
\hline
 \multicolumn{1}{|c|}{$\mnod$}
 & \multicolumn{4}{c|}{$(f\sqrt{M})_{PS}/(f\sqrt{M})_V$} \\
\hline 
1.71 & 0.87(2)   & 0.87(4)   & 0.88(9) & 0.87(2) \\
2.0  & 0.890(14) & 0.887(17) & 0.88(4) & 0.889(11) \\
4.0  & 0.951(14) & 0.93(3)   & 0.90(8) & 0.94(2) \\
8.0  & 0.983(9)  & 0.974(15) & 0.96(4) & 0.978(9) \\
\hline
\end{tabular}
\end{center}
\caption{Ratio of axial and vector matrix elements from Run A and
Run B, in lattice units.}
\label{table:ratio}
\end{table}
\begin{figure}[pthb]         
\vspace{-1cm}
\centerline{\epsfxsize=8cm
\hspace{0.6cm}\epsfbox{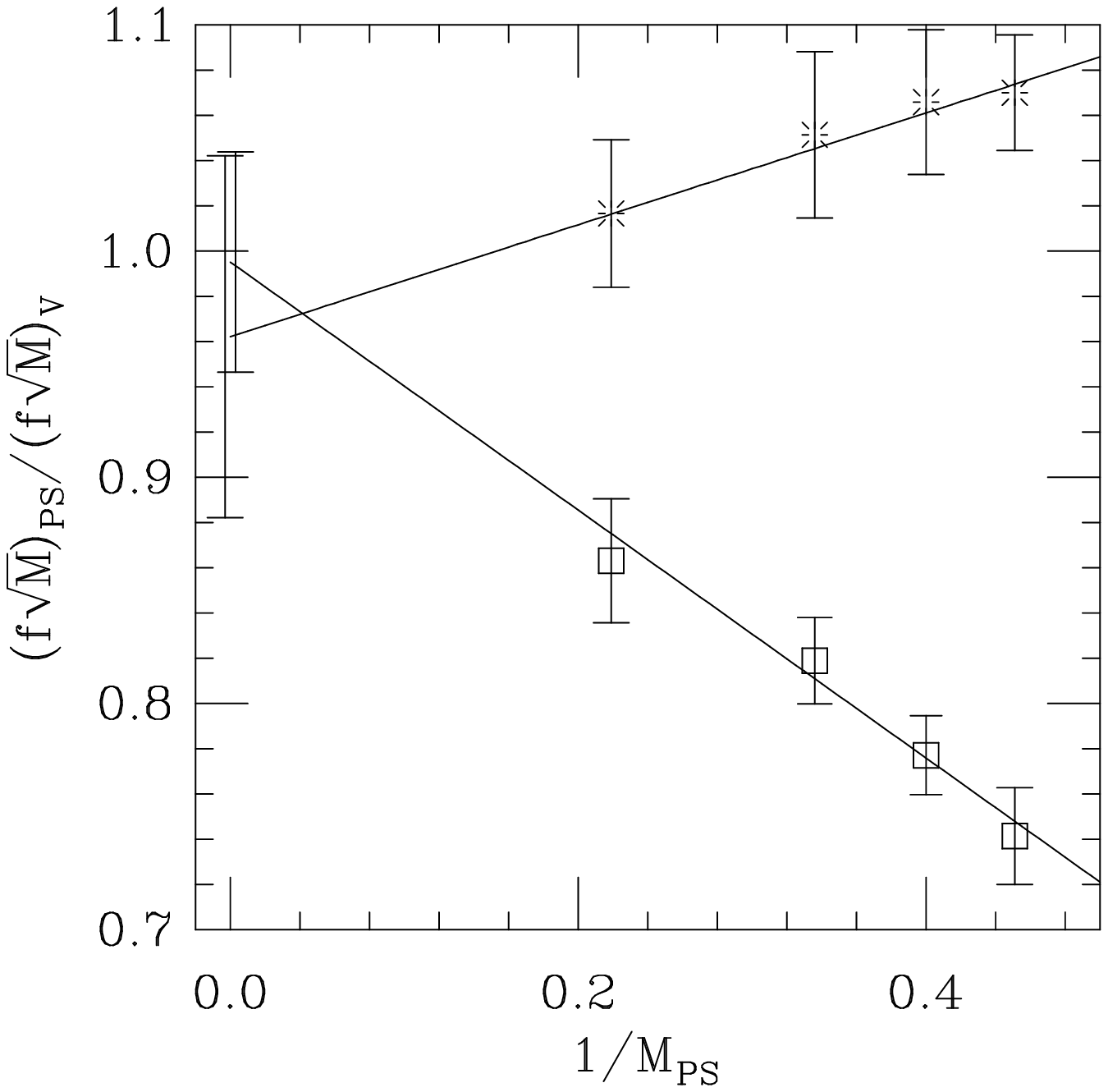}
\epsfxsize=8cm
\hspace{-0.4cm}\epsfbox{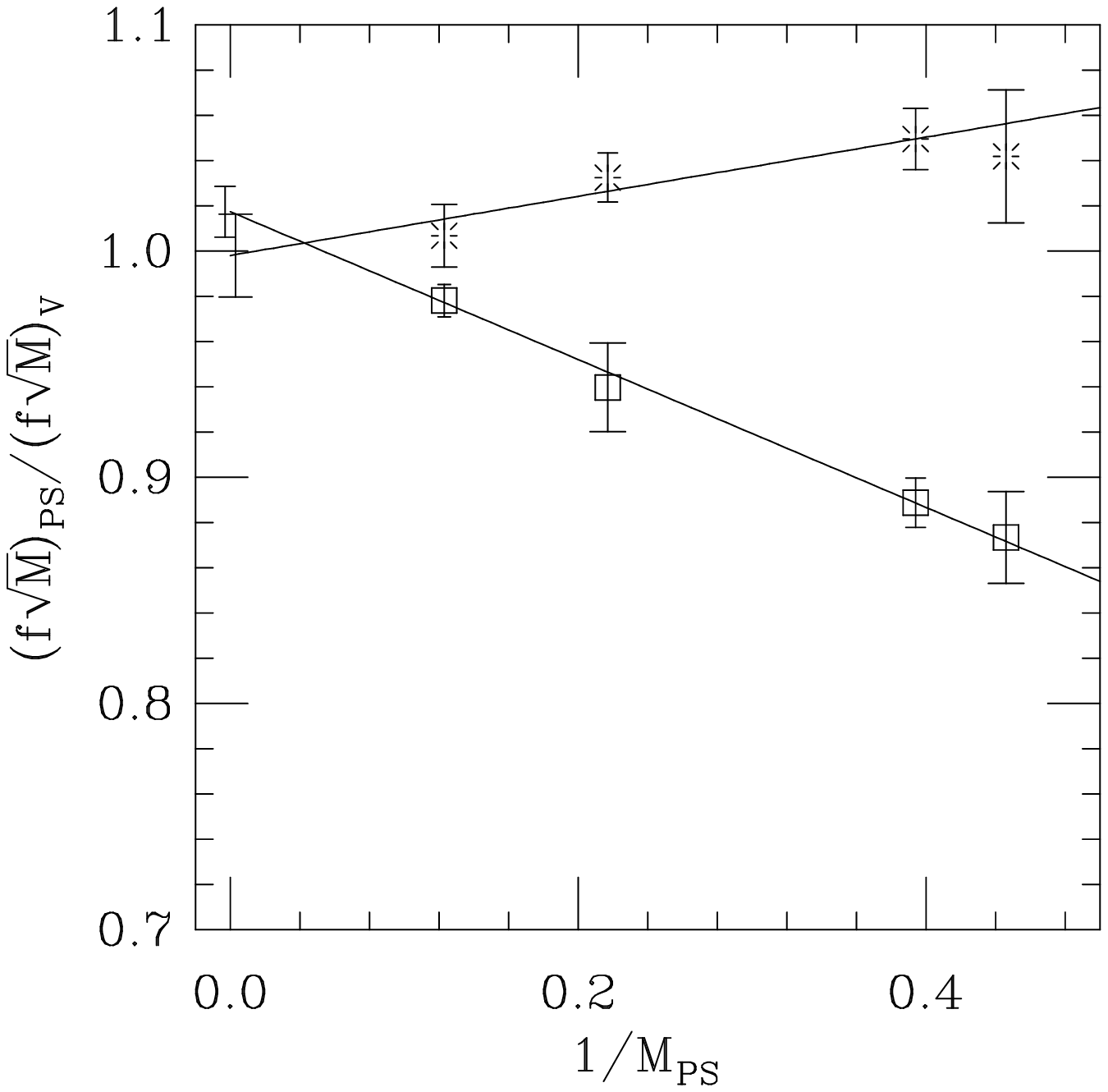}
}
\caption{Ratio of unrenormalised axial and vector matrix elements at
$\kappa = \kappa_{\rm s}$ from 
Run A (left) and Run B (right). Bursts refer to results without the current 
correction 
included, squares to results with the current correction. The lines denote 
correlated fits to the ratios.}
\label{fig:ratio}
\end{figure}
\begin{table}
\begin{center}
\begin{tabular}{|l|l|c|l|c|c|}
\hline
\multicolumn{3}{|c|}{Run A} &
\multicolumn{3}{c|}{Run B} \\
\hline
\multicolumn{1}{|c|}{$a_1$} &
\multicolumn{1}{c|}{$a_2$} &
\multicolumn{1}{c|}{$Q$} &
\multicolumn{1}{c|}{$a_1$} &
\multicolumn{1}{c|}{$a_2$} &
\multicolumn{1}{c|}{$Q$} \\
\hline
\multicolumn{6}{|c|}{$R^{uncorr}$} \\
\hline
0.96(8) &  0.25(19) & 0.97 & 0.998(18) & 0.13(7) & 0.69\\
1 (fixed) & 0.16(5) & 0.92 & 1 (fixed) & 0.12(3) & 0.86 \\
\hline
\multicolumn{6}{|c|}{$R$} \\
\hline
1.0(5)    & $-0.55(13)$ & 0.74 &  1.017(11) & $-0.33(5)$ & 0.93\\
1 (fixed) & $-0.56(3)$  & 0.90 &  1 (fixed) & $-0.27(2)$  & 0.42\\
\hline
\end{tabular}
\end{center}
\caption{Results  from correlated fits of the ratio of axial and vector decay
matrix elements, in lattice units, 
to the function $a_1 + a_2/M_{PS}$. $\kappa$ = $\kappa_{\rm s}$.
$R^{uncorr}$ is the 
ratio of the uncorrected matrix elements, $R$ has the current corrections 
included. }
\label{tab:fit_R}
\end{table}
The correlated ratios of axial and vector matrix elements from both
runs are listed in table~\ref{table:ratio}. The ratios can for
both runs be fit to a linear function in $1/M_{PS}$. The fit results are
presented in Table~\ref{tab:fit_R}. As expected, we find the extrapolation to
the infinite mass limit to be in good agreement with one. We also perform fits
with the value of the ratio fixed to one at infinite mass. In some cases these fits 
are slightly worse (see Table~\ref{tab:fit_R}); however the slopes 
from both fit methods agree with  each other. We find the slope of 
the uncorrected ratio and thus  $G_{hyp}$ to be slightly larger for Run A than
for Run B, but the difference is not statistically significant. We note that the
uncorrected ratios are independent of the light quark mass, which we also 
found for the $B^\ast - B$ hyperfine splittings~\cite{us_quenched}. The 
absolute value of the slope of the ratio $R$ 
turns also out to be larger in Run A than in Run B. Since the sign of the 
combination $8G_{hyp}+2G_{corr}/3$ is opposite to $G_{hyp}$, this means that
the current correction in Run A is considerably larger than in Run B. Another, more
precise, way to determine $G_{corr}$ is from the ratio of the current corrections to
the uncorrected current~\cite{Sara}. From Eq~(\ref{eq:neubert}) it follows that
\beq
G_{corr} = \lim_{\mnod \rightarrow \infty} \rho(\mnod) \equiv 
\lim_{\mnod \rightarrow \infty} \frac{2\mnod \delta(f\sqrt{M})_{PS}}
{(f\sqrt{M})_{PS}^{uncorr}}. \label{eq:gcorr}
\eeq
\begin{table}[thpb]
\begin{center}
\begin{tabular}{|l|l|l|l|l|}
\hline
\multicolumn{1}{|c}{} &
\multicolumn{2}{|c|}{Run A} &
\multicolumn{2}{|c|}{Run B} \\
\hline
\multicolumn{1}{|c|}{$\mnod$} &
\multicolumn{1}{c|}{$\kappa = \kappa_{\rm crit}$} &
\multicolumn{1}{c|}{$\kappa = \kappa_{\rm s}$} &
\multicolumn{1}{c|}{$\kappa = \kappa_{\rm crit}$} &
\multicolumn{1}{c|}{$\kappa = \kappa_{\rm s}$} \\
\hline
 1.71 & 0.2030(16) & 0.1978(3) & 0.1494(18) & 0.1403(5) \\
 2.0  & 0.1747(10) & 0.1707(3) & 0.1265(15) & 0.1227(3) \\
 2.5  & 0.1410(8)  & 0.1385(3) &            &           \\ 
 4.0  & 0.08988(10)& 0.0884(3) & 0.0688(7) & 0.06642(17)\\
 8.0  &            &           & 0.0353(3) & 0.03409(6) \\
\hline
\end{tabular}
\end{center}
\caption{Ratio $|\delta(f\protect\sqrt{M})_{PS}/
(f\protect\sqrt{M})_{PS}^{uncorr}|$ of the axial current correction 
to the uncorrected matrix element, in lattice units,
at the chirally extrapolated and the strange quark mass.}
\label{table:gcorr}
\end{table}
\begin{figure}[bhtp]         
\vspace{-1cm}
\centerline{\epsfxsize=8cm
\epsfbox{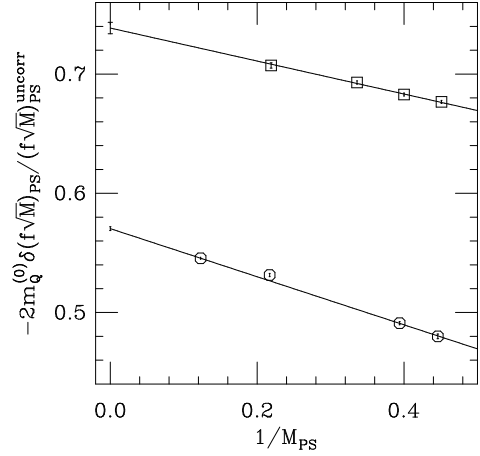}
}
\caption{Ratios of unrenormalised current corrections at $\kappa = \kappa_s$
plotted as a function of the inverse heavy meson mass.
Squares denote results from Run A, circles results from Run B; the lines are
correlated fits to the function $a_1 + a_2/M_{PS}$. }
\label{fig:gcorr}
\end{figure}
We fit $\rho(\mnod)$  from Run A to a linear function in 
$1/M_{PS}$ and find $G_{corr} = -0.739(5)$. If we include all values for $\rho$ 
from Run B  in a linear fit, we obtain a bad fit ($Q = 0.01$). The situation
improves with a quadratic fit ($Q=0.16$), but we obtain a better result from 
 a linear 
fit, omitting the point at $\mnod = 4.0$ from the fit. This gives $Q=0.44$, and for
$G_{corr}$ we obtain $-0.5703(16)$. Both fits are shown in Fig.~\ref{fig:gcorr}.
In summary, we find  each of the three different $1/M$ corrections to $f\sqrt{M}$, as well
as the total (compare Tables~\ref{table:fsqrtM} and~\ref{tab:fsqrtMtadpole1}, and
Tables~\ref{table:deltafsqrtM_A} and~\ref{table:deltafsqrtM_B}), 
in Run A to be larger  than in Run B.
\subsubsection{\boldmath$f_{B_s}/f_{B_d}$}
The ratio  $f_{B_s}/f_{B_d}$ can be used for an extraction of the ratio of
CKM matrix elements $|V_{ts}/V_{td}|$~\cite{CKM}. HQET predicts that it
is up to corrections $O(m_s-m_d)/m_Q$ independent of the heavy quark mass.
The renormalisation constants are expected to be very weakly dependent on 
the light quark mass. We approximate the physical ratio  $f_{B_s}/f_{B_d}$
with the ratio of the unrenormalised lattice matrix elements:
\begin{equation}
\frac{f_{B_s}}{f_{B_d}} \simeq \frac{(f\sqrt{M})_{B_s}}{(f\sqrt{M})_{B_d}}
\label{eq:ratio}
\end{equation}
Note that the renormalisation constants used in the following subsection 
assume massless light quarks.  The results for $f_{B_s}/f_{B_d}$ are listed in 
Table~\ref{table:res_phys}. As expected, they are within errors independent of 
the heavy quark mass. We find that the ratio is larger for Run A, although not 
significantly. This disagreement might be a reflection of the different 
discretisation effects in both runs.
\begin{table}[htbp]
\begin{center}
\begin{tabular}{|r|r|r|}
\hline
\multicolumn{1}{|c}{$m_Q^{(0)}$} &
\multicolumn{1}{|c|}{Run A} &
\multicolumn{1}{c|}{Run B} \\
\hline
$1.71$ & $1.23(18)$ &   1.13(8) \\
 $2.0$ & $1.26(15)$ &   1.17(7) \\
 $2.5$ & $1.32(12)$ &    \\
 $4.0$ & $ 1.25(25)$ &  1.18(8) \\
 $8.0$ &             &  1.17(9)  \\
 static& $1.13(6)$ &    1.10(6)  \\
\hline
\end{tabular}
\end{center}
\caption{Ratios of decay constants $f_{B_s}/f_{B_d}$.}
\label{table:res_phys}
\end{table}
\subsection{\label{sec:ren} Renormalised matrix elements}
The results presented in the previous subsection are at tree level, i.e. they do not 
include the renormalisation constants which are required to match between the matrix 
elements in the effective theory on the lattice and the matrix elements in 
full QCD in the continuum. 

In NRQCD, the operators that contribute to the heavy-light
current mix under renormalisation. The matrix elements of the bare NRQCD current operators
on the lattice, $J^{(i)}_{latt}$, the renormalised NRQCD current  operators, 
$J^{(i)}_{ren}$, and the heavy-light current in full QCD, $J_{QCD}$, are thus 
related by:
\beq
\langle J_{QCD}\rangle = \sum_i \eta_i \langle J^{(i)}_{ren}\rangle
= \sum_i \eta_i \sum_j Z_{ij}^{-1}\langle J^{(j)}_{latt} \rangle. \label{eq:ren}
\eeq
At $O(1/\mnod)$, three operators contribute to the axial vector current:
\beq
J_{latt}^{(0)} = q_{34}^\dagger Q,\;\;
J_{latt}^{(1)} =  -i\frac{1}{2\mnod}q_{12}^\dagger\vec{\sigma}\cdot\vec{D}Q,\;\;
J_{latt}^{(2)} =  i\frac{1}{2\mnod}q_{12}^\dagger\stackrel{\leftarrow}{D}\cdot
\vec{\sigma}Q.
\eeq
$J^{(2)}_{latt}$ does not contribute at tree level. Although we did not simulate 
$J^{(2)}_{latt}$, we know its matrix element exactly, since translation invariance 
implies that at zero momentum $\langle J^{(2)}_{latt}\rangle = \langle J^{(1)}_{latt}
\rangle$.

Expanding the renormalisation constants $\eta_i$ and $Z_{ij}$ through $O(\alpha)$, 
Eq.~(\ref{eq:ren}) becomes
\begin{eqnarray}
\langle A^0_{QCD}\rangle & = & \left[1 + \alpha(B_0 - \zeta_{00}
- \zeta_{10} - \frac{1}{2}(C_Q + C_q))\right] \langle J^{(0)}_{latt}\rangle \nonumber\\
&+ & \left[1 + \alpha(B_1 - \zeta_{01}
- \zeta_{11} - \frac{1}{2}(C_Q + C_q))\right] \langle J^{(1)}_{latt}\rangle \nonumber \\
& + &\left[\alpha(B_2 - \zeta_{02}
- \zeta_{12}) \right] \langle J^{(2)}_{latt}\rangle. \label{eq:ZA}
\end{eqnarray}
The coefficients $B_i$ originate from the vertex and wave function renormalisation in
the continuum, the $\zeta_{ij}$ denote the vertex renormalisations on the lattice, and
$C_Q$ and $C_q$ the heavy and the light wave function renormalisation on the lattice,
respectively. $\alpha$ is calculated using the two-loop formula for 
$\alpha_V$~\cite{lepage93}.  We assume a reasonable choice for the scale for $\alpha_V$
lies between  $q^\ast = 1/a$ and $q^\ast = \pi/a$. 
 The contribution of the
rotation to the heavy-light vertex with NRQCD heavy quarks has not been calculated, so 
we cannot include the renormalisation constants to Run A. 
\begin{table}
\begin{center}
\begin{tabular}{|l|c|c|c|c|}
\hline
\multicolumn{1}{|c}{} &
\multicolumn{2}{|c|}{$\kappa_{\rm crit}$} &
\multicolumn{2}{c|}{$\kappa_{\rm s}$} \\
\multicolumn{1}{|c|}{$a\mnod$} &
\multicolumn{1}{c|}{$aq^\ast=1$} &
\multicolumn{1}{c|}{$aq^\ast=\pi$} &
\multicolumn{1}{c|}{$aq^\ast=1$} &
\multicolumn{1}{c|}{$aq^\ast=\pi$} \\
\hline
1.71     & 0.143(14) & 0.149(14) & 0.160(4) & 0.166(4) \\
2.0      & 0.141(11) & 0.149(12) & 0.161(3) & 0.170(3) \\
4.0      & 0.146(13) & 0.163(14) & 0.167(3) & 0.186(4) \\
8.0      & 0.146(15) & 0.172(16) & 0.167(5) & 0.195(5) \\
$\infty$ & 0.146(11) & 0.186(12) & 0.163(4) & 0.208(4) \\
\hline
\end{tabular}    
\end{center}
\caption{Renormalised $a^{3/2}(f\protect\sqrt{M})_{PS}$ from Run B at strange and 
chirally extrapolated light quark masses.}
\label{tab:ren}
\end{table}
\begin{figure}[pthb]         
\vspace{-2cm}
\centerline{\epsfxsize=13cm
\epsfbox{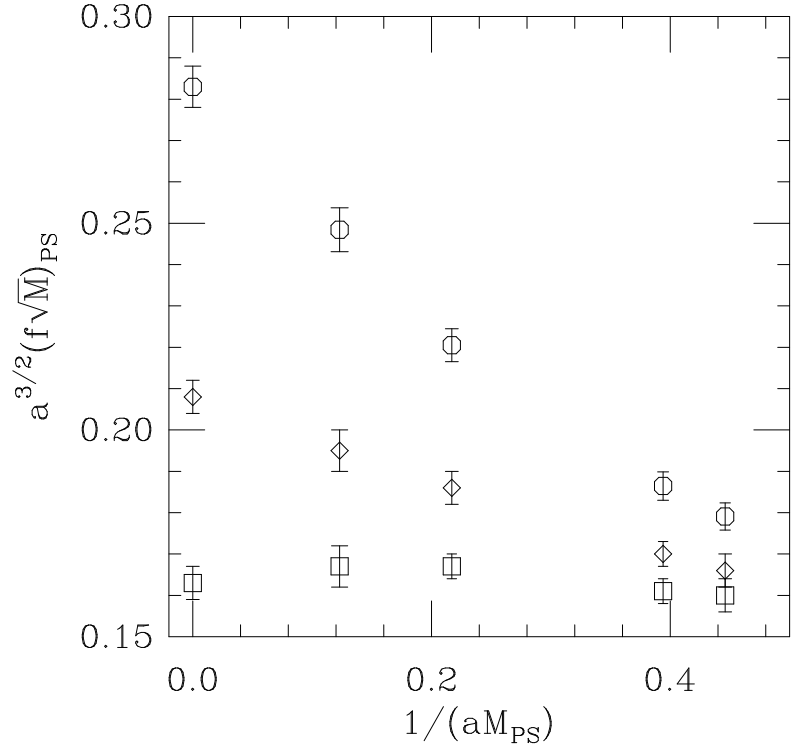}
}
\caption{Matrix elements $a^{3/2}(f\protect\sqrt{M})_{PS}$ from Run B
at $\kappa_{\rm s}$. The circles 
denote unrenormalised matrix elements; the squares denote renormalised
results using $aq^\ast = 1$, and the diamonds, using $aq^\ast = \pi$.}
\label{fig:ren}
\end{figure}
The renormalised matrix elements for Run B have been calculated using $B_i$, 
$\zeta_{ij}$ and $C_Q$ from Ref.~\cite{junko}, and $C_q$ from 
Ref.~\cite{goeckeler}. The results for chirally extrapolated and strange
light quark masses are given in Table~\ref{tab:ren}. Our renormalised
matrix elements include an $O(\alpha a)$ discretisation correction 
to $\langle J^{(0)}_{latt}\rangle$. The origin of this correction and its 
relation to Eq.~(\ref{eq:ZA}) will be discussed in future 
publications~\cite{junko,tsukuba,our_letter}. In 
Fig.~\ref{fig:ren} we compare the renormalised $a^{3/2}(f\sqrt{M})_{PS}$ for both 
values of $q^\ast$  with the unrenormalised matrix element. After renormalisation,
the slope for large masses is remarkably smaller than before.
For $aq^\ast$ = 1, it is within errors in agreement with zero.
For $aq^\ast$ = $\pi$, the $M$ dependence of the matrix elements is approximately
linear, and  we estimate the relative slope (the slope divided by the value at 
infinite mass) to be of the order of $-1$ GeV. The latter value is roughly in agreement 
with previous calculations; e.g. in Refs.~\cite{kappac,onogi,fB_MILC} the relative 
slope is $\sim\! -1$ GeV. Since the results at heavy masses 
vary considerably depending on $q^\ast$, we do not want to 
make a more quantitative statement here.

For $f_B$, we use the result at $a\mnod = 2.0$, 
the point in our simulation whose mass is closest to the actual $B$ meson (see 
Table~\ref{tab:shifts}):
\begin{eqnarray}
q^\ast = 1/a:    & f_B  = 0.174(28)(26)(16)\mbox{ GeV} \nonumber \\
                 & f_{B_s}  = 0.198(8)(30)(17)\mbox{ GeV} \nonumber \\                 
q^\ast = \pi/a: & f_B  = 0.183(32)(28)(16)\mbox{ GeV} \nonumber \\
                & f_{B_s} = 0.209(8)(32)(17)\mbox{ GeV} \label{eq:fb1}
\end{eqnarray}
This is to be compared with the tree level result, $f_B$
  = 0.195(22)(29)(7) GeV, and $f_{B_s}$ = 0.229(8)(35)(8).
The first error bar is the statistical error,
inflated by a factor of 2 to take the fitting uncertainty of $\sim\! 1\sigma$ 
(see Sec.~\ref{sec:runb}) into account. For $f_B$, the statistical error gets in
addition enlarged due to the chiral extrapolation. The second error bar stems from 
the uncertainty 
in the determination of $a$. The third one consists of the estimated error of the 
perturbative calculation and due to neglected orders in the $1/M$ expansion, and 
was determined as follows: 
The uncertainty from the choice of $q^\ast$ is 
$5\%$, which can be used as an estimate of the uncertainty from higher orders in
perturbation theory. Another estimate of this uncertainty, $\delta_2$, can be 
obtained from the relation $\alpha^2\delta_2 = (\alpha\delta_1)^2$, where $\delta_1$ 
is the one-loop contribution. Renormalisation decreases the decay constant 
for $q^\ast$ between $\pi/a$ and $1/a$ by $9- 10\%$ from the bare value. Squaring
this yields a contribution of $\sim\! 1\%$ from higher orders.  
The numerical errors on the integrals in the perturbative calculation are
estimated to be $\sim\! 2-3\%$ and propagate to the final result after being
multiplied by $\alpha$.
Another source of error are higher order contributions in the $1/\mnod$
expansion. A  calculation that includes also the $O(1/(\mnod)^2)$ 
corrections~\cite{StLouis,our_letter} indicates  that these corrections are in the 
region of the $B$ of the order of $\sim\! 3-4\%$. Adding this to the errors from
the perturbative calculation,  we quote a systematic
error of $9\%$ of the average from both $q^\ast$ on our results in Eq.~(\ref{eq:fb1}),
which is represented by the third error bar. For the unrenormalised matrix elements, 
the third error bar consists only of the $O(1/(\mnod)^2)$ correction. 
The largest errors on our results on $f_B$ come therefore  from the statistical and
fitting error, which is magnified due to chiral extrapolation, and the uncertainty in 
$a$. Note that at the $B$, the error from
the uncertainty in $q^\ast$ is much smaller than for the higher masses. The quenching 
error is partly reflected in the error we quote for the lattice spacing. There are
indications that unquenching leads to a larger value for $f_B$~\cite{fB_MILC,saraprep}.

For the static matrix element from Run B we obtain:
\begin{eqnarray}
q^\ast = 1/a: &
f_B^{\infty}  = 0.180(14)(27)(50)\mbox{ GeV} \nonumber \\
q^\ast = \pi/a:  & f_B^\infty  = 
0.229(15)(34)(50)\mbox{ GeV}, 
\end{eqnarray}
while the bare result is $a^{3/2}(f\sqrt{M})^\infty = 0.254(15)$. An estimate of 
$aq^\ast = 2.18$ has been given in Ref.~\cite{hernandez}
 for static heavy and Wilson light quarks.
In the static case the one-loop contribution to $Z_A$ is larger than for NRQCD around
the $b$ quark mass, thus also the variation with $q^\ast$ is larger ($24\%$). Here,
the dominant error appears to originate from higher orders in perturbation theory.


For the static matrix element, the renormalisation constant has been 
calculated also for rotated clover fermions~\cite{borrelli,junko}; however the
$O(\alpha a)$ discretisation correction has not been determined for this case. Without
this discretisation correction, we obtain for the static $f_B$ from Run A:
\begin{eqnarray}
q^\ast = 1/a:   & a^{3/2}(f\sqrt{M})^\infty = 0.154(10) & 
f_B^{\infty}  = 0.190(12)(21)\mbox{ GeV} \nonumber \\
q^\ast = \pi/a: & a^{3/2}(f\sqrt{M})^\infty = 0.202(14) & f_B^\infty
  = 0.249(17)(27)\mbox{ GeV}, \label{eq:untad}
\end{eqnarray}
compared to the bare result $a^{3/2}(f\sqrt{M})^\infty = 0.281(19)$. Note also that
the result in Eq.~(\ref{eq:untad}) is not tadpole-improved.
\section{Summary and Conclusions}
We report on a study of quenched heavy-light decay constants with non-relativistic heavy 
quarks in a mass range around the $b$ quark and heavier. Both the
NRQCD  Lagrangian and heavy-light current are correct through $O(1/m_Q^{(0)})$. 
We performed two simulations, one
uses clover light quarks with tree-level clover coefficient (Run A), the other 
uses tadpole-improved light quarks (Run B). 

We investigated the $1/M$ behaviour of the unrenormalised 
decay matrix elements $f\sqrt{M}$, $M$ being the heavy-light meson mass. 
In the mass region of the $B$ meson, the correction to the 
static limit is large (for Run A $\sim \!50\%$, for Run B $\sim \!35-40\%$), before
renormalisation constants are included. 
We disentangle the various $1/M$ corrections to the
decay constants, and compare their size  between Run A and Run B. 
The differences are small for the contributions of the 
hyperfine and  kinetic term in the action, but sizeable for the current correction 
matrix element, at least at tree level.
At the $B$, we find for the bare axial matrix element a difference  of 
18\% ($1-2\sigma$) for chirally extrapolated light quarks and 10\% ($2-3\sigma$) for strange
light quarks. We expect this difference to be partly caused by a reduction of the 
$O(\alpha a)$ errors in Run B due to tadpole improvement. The fact that we use a rotation 
with a derivative operator in Run A and a normalisation with $\sqrt{1-6\tilde{\kappa}}$
in Run B  also introduces a difference in the  discretisation effects between the 
two runs. 

We compare $f\sqrt{M}$ using NRQCD with results using  clover ($c_{SW} = 1$) heavy quarks 
generated by UKQCD. Renormalisation constants are not included. 
In the region of the $b$ quark both methods agree within errors. However, they 
behave quite differently at large masses, such that the clover results
cannot be made to  extrapolate to the static limit. 

We calculated the renormalisation constant $Z_A$  for NRQCD
in one-loop perturbation theory, taking into account the mixing between the
current operators. For the first time, we present renormalised 
pseudoscalar decay constants, and a value for $f_B$ from NRQCD where all the 
matching factors through  $O(\alpha/M)$ are included. The bare matrix
elements show a larger slope in $1/M$ than the results from calculations with
relativistic heavy quarks, but the heavy mass dependence
of the renormalised matrix elements $f\sqrt{M}$ is much milder than before 
renormalisation.
\subsection*{Acknowledgements}
We are grateful to the UKQCD Collaboration for allowing us to use their gauge
configurations and light propagators. We would like to thank P. Lepage 
for useful discussions. This work was supported by SHEFC, PPARC,
the U.S. DOE (contract DE-FG02-91ER40690), the NATO under grant
number CRG 941259, and the EU grant CHRX-CT92-0051.  A. A. would like
to thank the Graduate School of the Ohio State 
University for a University Postdoctoral Fellowship. We thank the 
Edinburgh Parallel Computing Centre for computer time on their CM-200, and
the Ohio Supercomputer Center  for time on their CRAY Y-MP. 

\end{document}